\documentclass[english,letter,12pt,twosided]{article}
\usepackage{UF_FRED_paper_style}

\usepackage{enumitem}
\usepackage{tikz}
\usepackage{mathrsfs}
\usepackage{threeparttable}
\usepackage{amsmath}
\usepackage{booktabs}
\usepackage{bookmark}
\usepackage{caption}
\usepackage{graphicx}
\usepackage[T1]{fontenc}
\usepackage[utf8]{inputenc}
\usepackage{longtable}

\usepackage{amsthm}

\theoremstyle{plain}

\theoremstyle{definition}

\theoremstyle{remark}

\usepackage{datatool}
\usepackage{pdflscape}

\DTLloaddb[keys={key,value}]{stats}{paper_stats.csv}
\newcommand{\getstat}[1]{%
  \DTLfetch{stats}{key}{#1}{value}\unskip
}

\definecolor{PennBlue}{RGB}{001,031,091}
\definecolor{PennRed}{RGB}{153,0,0}
\hypersetup{
pdfborder = {0 0 0},
    colorlinks,
    citecolor=PennRed,
    filecolor=PennRed,
    linkcolor=PennRed,
    urlcolor=PennRed}

\title{\textbf{The Rise of Negative Earnings and Demand Shifting Investment}}

\author{Jacob Toner Gosselin\thanks{We thank Lawrence Christiano, Martin Eichenbaum, Benjamin Jones, Kiminori Matsuyama, Sara Moreira, Matthew Rognlie, Kunal Sangani and Alireza Tahbaz-Salehi for their guidance. We also thank Aaron Amburgey, José Luis Lara, Ramya Raghanavan, and the participants of the Northwestern Macro Lunch Seminar for their helpful comments.} \\ \href{mailto:jacob.gosselin@u.northwestern.edu}{Northwestern University} 
\and Dalton Rongxuan Zhang \\ \href{r.zhang@u.northwestern.edu}{Northwestern University}
    }

\date{\today}

\begin{document}
\begin{filecontents*}{paper_stats.csv}
"key","value"
"neg_earnings_1980",10.81
"neg_earnings_2019",28.04
"neg_spell_1980",1.93
"neg_spell_2019",4.08
"avg_earnings_percchange",282.75
"med_earnings_percchange",275.07
"avg_earnings_factorchange",3.83
"med_earnings_factorchange",3.75
"num_obs_analysisdata",200120
"num_firms_analysisdata",19141
"sd_sales_percchange",129.15
"sd_ebitda_percchange",188.76
"cogs_change_all",-9.97
"capex_change_all",-1.43
"sga_change_all",5.13
"cogs_change_neg",-19.07
"capex_change_neg",-2.65
"sga_change_neg",27.6
"rd_change_neg",11.68
"neg_spell_1980_model",1.6
"neg_spell_2019_model",3.16
"sd_earnings_percchange_model",21.39
"sd_sales_percchange_model",50.34
"adv_ratio_1980_model",0.11
"adv_ratio_2019_model",0.22
"inv_ratio_change_model",18.37
"gdp_pct_change",-9.1
"c_pct_change",-16
"i_pct_change",-7
"a_pct_change",48.4
"sales_wtd_z_pct_change",-4.9
\end{filecontents*}

\captionsetup{justification=centering, singlelinecheck=false}
\vfill
{\setstretch{.8}
\maketitle
\begin{abstract}
We document the rise of negative earnings between 1980 and 2019: a secular increase in the percent of firms reporting losses, both among public firms and in the broader universe of US corporations, and a secular increase in the persistence of losses year-to-year among public firms. This rise has occurred alongside a spreading of the sales and earnings distribution and a recomposition of firm spending away from production costs and traditional investment and towards selling, general and administrative expenses. We rationalize these phenomena with a model of heterogenous firms engaging in \emph{supply and demand shifting investment}. Our model includes a \emph{scale elasticity of demand} determining the relationship between the intensive margin of demand (demand per customer) and the extensive margin of demand (number of customers). We are able to quantitatively match the rise in reported losses and qualitatively match (1) the increased persistence of losses, (2) the spreading of the sales and earning distribution and (3) the recomposition of firm spending with this parameter as the single driver of changes across steady state equilibria. The rise in the scale elasticity associated with the increase in reported losses has non-trivial aggregate implications: in our model it lowers GDP by \getstat{gdp_pct_change}\% by reallocating labor away from goods and capital production and reallocating demand away from productive firms.
\end{abstract}
}
\vfill


\section*{Introduction}

\begin{figure}[!h]
    \centering
    \includegraphics[width=\textwidth]{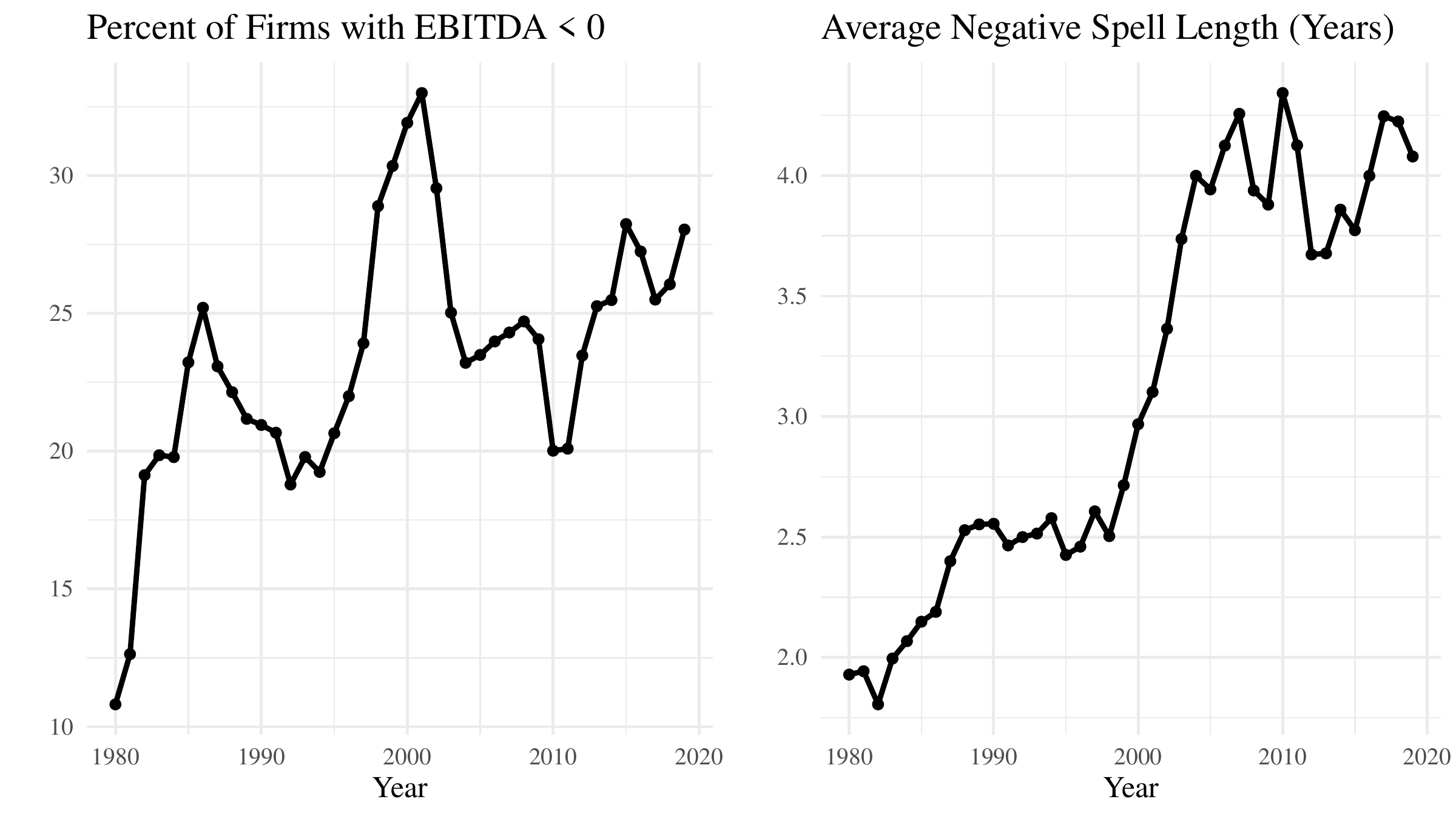}
    \caption{The Rise of Negative Earnings}
    \label{fig:neg_earnings_and_spell_over_time}
     \scriptsize\raggedright{\textit{The figure plots the percent of Compustat firms reporting losses for each year (left panel) and the average number of consecutive years Compustat firms with losses have reported losses (right panel) between 1980 and 2019, with earnings defined as EBITDA.}}
\end{figure}

A large empirical literature in economics has emerged documenting facts that fall under a common theme: US businesses behave differently today than they did fifty years ago. We aim to add to this set of facts the \emph{rise of negative earnings}: as shown in Figure~\ref{fig:neg_earnings_and_spell_over_time}, the percentage of public firms reporting losses in a given year rose from \getstat{neg_earnings_1980}\% in 1980 to \getstat{neg_earnings_2019}\% in 2019, while the average negative earnings ``spell'' among such firms, the number of consecutive years a firm with no earnings has had no earnings, rose from \getstat{neg_spell_1980} years to \getstat{neg_spell_2019} years.

These changes are puzzling on two fronts. First, over the same period public firms' earnings rose overall: in real terms, mean (median) earnings rose by a factor of \getstat{avg_earnings_factorchange} (\getstat{med_earnings_factorchange}) between 1980 and 2019. Second, over the same period variable costs (COGS) and physical investment (CapEx) fell as a share of sales, for all firms and for firms with negative earnings. If firms are earning more overall and spending less of their revenue on production costs and traditional forward-looking investment, then why are more firms losing money for longer?

We rationalize this phenomena with a model of heterogenous firms engaging in \emph{supply and demand shifting investment}. Firms are able to move their future supply curves by investing in their physical capital stock, and move their future demand curves by investing in their customer base. Our model includes a \emph{scale elasticity of demand}, the elasticity of demand per customer with respect to customer base size. Movements in this parameter alone allow us able to perfectly match the rise in the percent of firms reporting losses observed in the data, and qualitatively match three other untargeted trends: (1) the increase in the persistence of losses, (2) the widening of the sales and earnings distribution, and (3) the increase in selling, general and administrative spending (SG\&A) relative to COGS and CapEx. We present suggestive evidence that the scale elasticity of demand has indeed risen between 1980 and 2019, and discuss the aggregate implications of such a rise.

To examine changes in the earnings of public firms, we use data from Compustat with traditional filters applied. Our results in Figure \ref{fig:neg_earnings_and_spell_over_time} define earnings as self-reported EBITDA among Compustat firms. To confirm this phenomena generalizes to the broader universe of US firms, we use public data from the IRS Statistics of Income to document that the rise in the percent of firms reporting losses is observable in statistics covering all active US corporations, excluding S-Corps. Moreover, among public firms the rise in negative earnings is (1) robust to alternative earnings definitions and (2) common across sectors.

In addition to the rise of negative earnings, we find that the real earnings distribution and real sales distribution both spread between 1980 and 2019, with the upper and lower quantiles both pulling away from the median. This spreading is how real earnings rose as negative earnings became more common. Moreover, we find that while COGS and CapEx fell as a share of sales for the median firm, SG\&A spending rose as a share of sales. This recomposition of spending is even more pronounced when looking at the median negative earning firm.

To rationalize these trends, we develop a model of heterogenous firms investing in supply and demand shifting capital. Our firms are heterogenous in their productivity a la \citet{hopenhaynJobTurnoverPolicy1993}. They accumulate physical capital through traditional investment, which shifts their supply curve, and grow their customer bases (accumulate ``customer capital'') through sales and marketing spending, an investment which shifts their demand curve. We define the steady state equilibrium of our model as the stationary distribution where agents optimize and markets clear. It serves as the object of our quantitative analysis. 

Unlike traditional investment, sales and marketing spending is (1) purely business stealing, since customers are in fixed supply in the aggregate, and (2) has returns that depend on the \emph{scale elasticity of demand}: the elasticity of demand per customer (the intensive margin of demand) with respect to customer base size (the extensive margin of demand). A larger scale elasticity of demand increases the returns to sales and marketing spending; since this investment is business stealing, firms over-invest more, not internalizing that their investment makes it more expensive to acquire customers. It also increases the optimal customer base size; since firms want to grow larger, they spend more on forward looking investment for longer. 

We solve for a steady state equilibrium corresponding to each year in our data, using the scale elasticity of demand as the sole driver of changes across equilibria. Because of the two mechanisms described above the percent of firms with negative earnings, average negative earning spell, sales and earnings distribution spread, and sales and marketing spending relative to other costs, \emph{are all monotonically increasing in the scale elasticity of demand across equilibria}. Thus, with this parameter alone we quantitatively match the rise in the percent of firms reporting losses observed in the data, while qualitatively matching (1) the rise of the average negative earning spell, (2) the widening of the sales and earnings distribution, and (3) the increase in SG\&A spending relative to COGS and CapEx. 

To examine whether the scale elasticity of demand has indeed risen over our sample period, we construct a proxy for customer capital by applying the perpetual inventory method to SG\&A spending \citep{eisfeldtOrganizationCapitalCrossSection2013}. Our model implies that, holding all other parameters fixed, the sales elasticity of customer capital is linear in the scale elasticity of demand. We exploit this reduced-form relationship, regressing logged sales on logged customer capital interacted with year (with relevant controls). Our year-specific estimates of the sales elasticity of customer capital correspond with a secular rise in the scale elasticity of demand. 

Finally, we consider the aggregate implications of a rise in the scale elasticity of demand. We find that the rise in the scale elasticity associated with the increase in reported losses observed in the data implies a \getstat{gdp_pct_change}\% loss in GDP in our model, due to a \getstat{c_pct_change}\% reduction in consumption. The fall in consumption is driven by changes in firms' investment choices \emph{reallocating labor} away from capital and goods production and \emph{reallocating demand} away from productive firms.

\paragraph{Related Literature}

As discussed in the introduction, our paper relates to a large literature on secular changes in US business dynamics, in particular work on the rise of intangible investment relative to traditional capital expenditures \citep{eisfeldtOrganizationCapitalCrossSection2013, crouzetEconomicsIntangibleCapital2022}, the role of sales and marketing spending as capital investment \citep{heInvestingCustomerCapital2024}, the spreading of the sales distribution and emergence of superstar firms \citep{autorFallLaborShare2020,kwon100YearsRising2024}, and the decline in public firm profitability as measured by operating cash flow over assets \citep{kahleUSPublicCorporation2017}. Most relevantly, the increase in the percentage of of public firms with losses and the persistence of losses among public firms were first documented in \citet{denisPersistentOperatingLosses2016}. Our contribution to this literature is two-fold. First, we document that the rise of negative earnings observed in Compustat data occurs in the broader universe of US corporations, using public data from the IRS Statistics of Income (SOI) on corporate tax returns. Second, we present a model with a single mechanism that can jointly account for the rise in the percent of firms reporting losses, the persistence of losses, the spreading of the sales and earnings distribution, and the rise of intangible sales and marketing investment relative to traditional CapEx.  

Our paper is also connected to a growing literature modeling sales and marketing spending by firms as an investment that shifts the extensive margin of demand for their product \citep{afrouziConcentrationMarketPower2023, arkolakisMarketPenetrationCosts2010, gourioCanIntangibleCapital2014, hubmerInvestmentDemandDynamic} as well as the broader literature modeling advertising \citep{bronnenbergMarketingInvestmentIntangible2022,greenwoodYouWillMacroeconomic2025}. Here we are most closely related to \citet{afrouziConcentrationMarketPower2023}, who present a model of customer capital as endogenous customer acquisition by monopolistically competitive firms. We extend their work by (1) allowing firms to invest in endogenous customer acquisition (demand shifting investment) \textbf{and} traditional physical capital (supply shifting investment), and (2) introducing scale effects in demand through the interdependence of demand per customer (the intensive margin of demand) and number of customers (the extensive margin of demand).

Finally, our discussion of the macro implications of changes in the scale elasticity of demand relates to the literature on misallocation across firms \citep{hsiehMisallocationManufacturingTFP2009a, afrouziConcentrationMarketPower2023, hubmerInvestmentDemandDynamic}. While much the previous literature has focused on the misallocation of factors of production across heterogenous firms, we focus on the misallocation of demand across firms, which in our model is endogenous to firms' investment decisions, as well as the misallocation of labor across good types (i.e. productions goods, capital goods, and advertisements).

The paper proceeds as follows: In Section \ref{sec:empirical} we discuss our empirical results, describing our data and the secular trends we seek to rationalize.  In Section \ref{sec:theory} we describe our model.  In Section \ref{sec:quantitative} we calibrate our model and highlight its ability to match the secular trends of we see empirically via changes in a single parameter: the scale elasticity of demand. In Section \ref{sec:aggregation} we discuss the effect of changes in the scale elasticity of demand on GDP and its sub-components: consumption, investment, and advertising.

\section{Motivating Trends}
\label{sec:empirical}

In this section we present our empirical results, which are a series of secular trends across firms. First, we document a rise in the incidence and duration of negative earnings as well as a spreading of the earnings and sales distribution. Second, we document a recomposition of firm spending away from COGS and CapEx and towards SG\&A.

\subsection{Data Construction}

Our main analysis uses annual firm-level data on US public firms in non-finance, non-utility sectors from Compustat. We restrict to consolidated, standard-format, USD-denominated annual reports with non-missing sales, COGS, and SG\&A, between 1980 and 2019. We also filter to firms with non-zero sales and COGS, and trim at the 1st and 99th percentile of the sales/COGS ratio. Our final sample has \getstat{num_obs_analysisdata} firm-year observations, covering \getstat{num_firms_analysisdata} unique firms.

Our main variables of interest are earnings, defined as EBITDA, as well as sales, COGS, SGA, R\&D, and CapEx. We separate R\&D spending from SG\&A as in \citet{petersIntangibleCapitalInvestmentq2017}. To examine the robustness of our negative earnings results, we include alternative earnings measures of net income (NI), pre-tax income (PI), and profits (Sales - COGS - SG\&A - CapEx). We report all variables in real terms, adjusting nominal values using the Bureau of Economic Analysis (BEA) GDP deflator. 

To measure negative spell lengths, we sum the number of consecutive years a firm has reported negative earnings, including the current year (e.g. if a firm reported positive EBITDA in 1979, negative EBITDA in 1980-1982, and positive EBITDA in 1983, then the negative spell lengths in each year would be $0, 1, 2, 3, 0$). To construct this measure we use Compustat data going back to 1961. Since our analysis begins at 1980, we cap all spell lengths at 19 years; in practice this only effects $0.036\%$ of observations.

We also use data from the IRS Statistics of Income on the proportion of corporate filings reporting income between 1980--2019, excluding S-Corps. All statistics used are publicly available via the \href{https://www.irs.gov/statistics/soi-tax-stats-corporation-income-tax-returns-complete-report-publication-16}{Corporation Complete Report}; they are aggregated in Table \ref{tab:irs_data}.

To calibrate model parameters, we estimate production functions using the methodology of \citet{ackerbergIdentificationPropertiesRecent2015} (ACF), and estimate mark-ups using the ratio estimator \citep{deloeckerMarkupsFirmLevelExport2012}.

Additional details on our data construction and production function estimation can be found in Appendix \ref{ap:empirical}.

\subsection{Empirical Results}
\label{subsec:empirical_results}

\paragraph{The Evolution of Firm Earnings}

Our main empirical result is the rise in negative earnings summarized in Figure~\ref{fig:neg_earnings_and_spell_over_time}: in our Compustat sample, the percentage of firms with $\text{EBITDA} < 0$ has risen from \getstat{neg_earnings_1980}\% to \getstat{neg_earnings_2019}\% between 1980 and 2019, while the average number of consecutive years a firm with $\text{EBITDA} < 0$ has reported negative EBITDA rose from \getstat{neg_spell_1980} to \getstat{neg_spell_2019}.

\begin{figure}[!h]
    \centering
    \includegraphics[width=\textwidth]{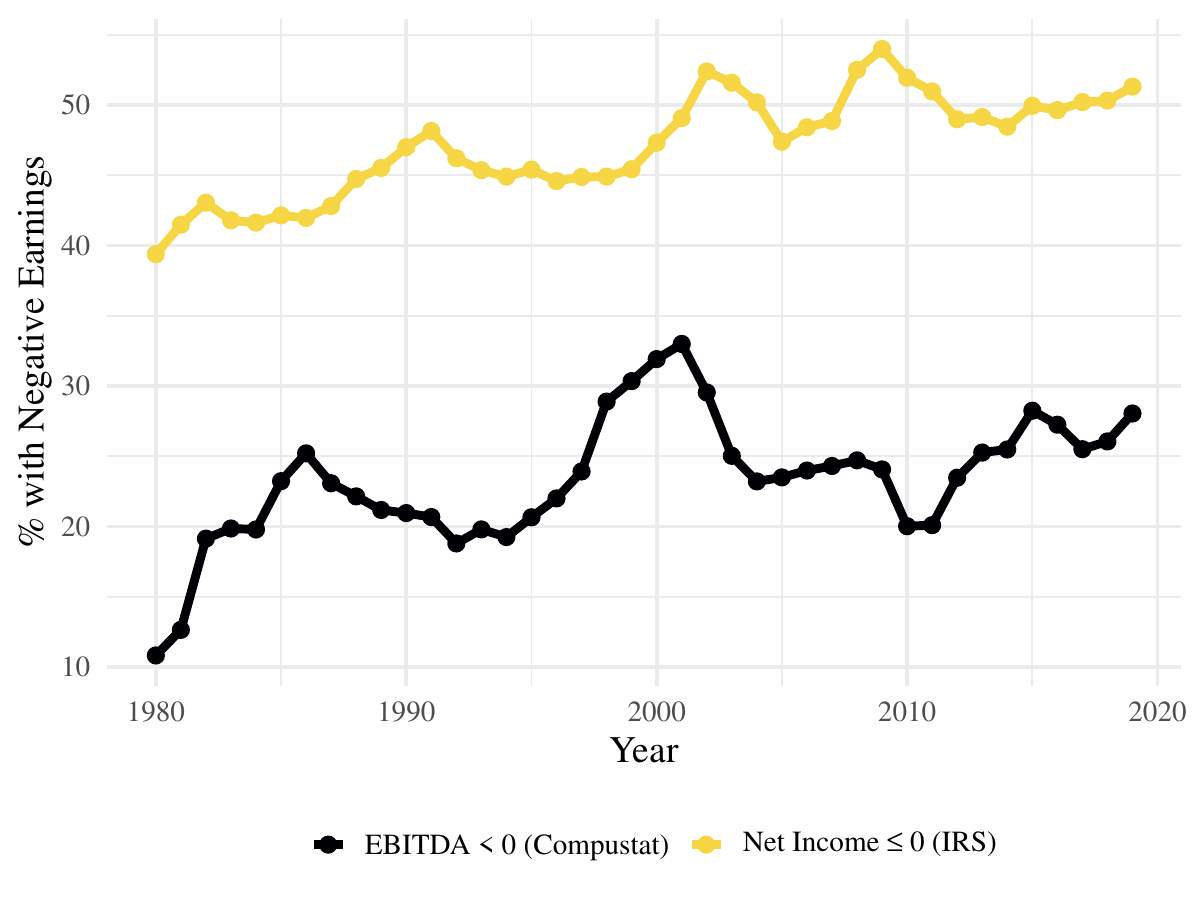}
    \caption{The Rise of Negative Earnings in Compustat and IRS Data}
    \label{fig:pct_negative_earnings_compustat_vs_irs}
    \scriptsize\raggedright{\textit{The figure plots (A) the percent of Compustat firms inour sample reporting EBITDA<0 and (B) the percent of active corporate returns (excluding S-Corps) with no income as reported in the IRS SOI, for each year between 1980-2019.}}
\end{figure}

Since Compustat data is not always representative of the broader universe of firms, it is worth considering whether this phenomena is present in data that includes private corporations. To that end, we use the number of corporate filings with income and the total number of corporate filings as reported in the IRS SOI \href{https://www.irs.gov/statistics/soi-tax-stats-corporation-income-tax-returns-complete-report-publication-16}{Corporation Complete Report} to generate an annual measure of the negative earnings incidence rate for the universe of active US Corporations, excluding S-Corps. As shown in Figure~\ref{fig:pct_negative_earnings_compustat_vs_irs}, this measure shows a qualitatively similar rise between 1980 and 2019. The IRS data is used to generate the figure is visible in Table \ref{tab:irs_data}.

To confirm that the increased persistence of losses we observe in the cross-section isn't driven by any single cohort, we calculate the average negative spell length for 5-year cohorts from 1962-2010, grouping firms by the first year they appear in Compustat data. For earnings defined as EBITDA and profits, we see a similar rise in average negative spell length across the cohorts, as shown in Figure \ref{fig:neg_spell_cohorts}. 

\begin{figure}[!h]
    \centering
    \includegraphics[width=0.95\textwidth]{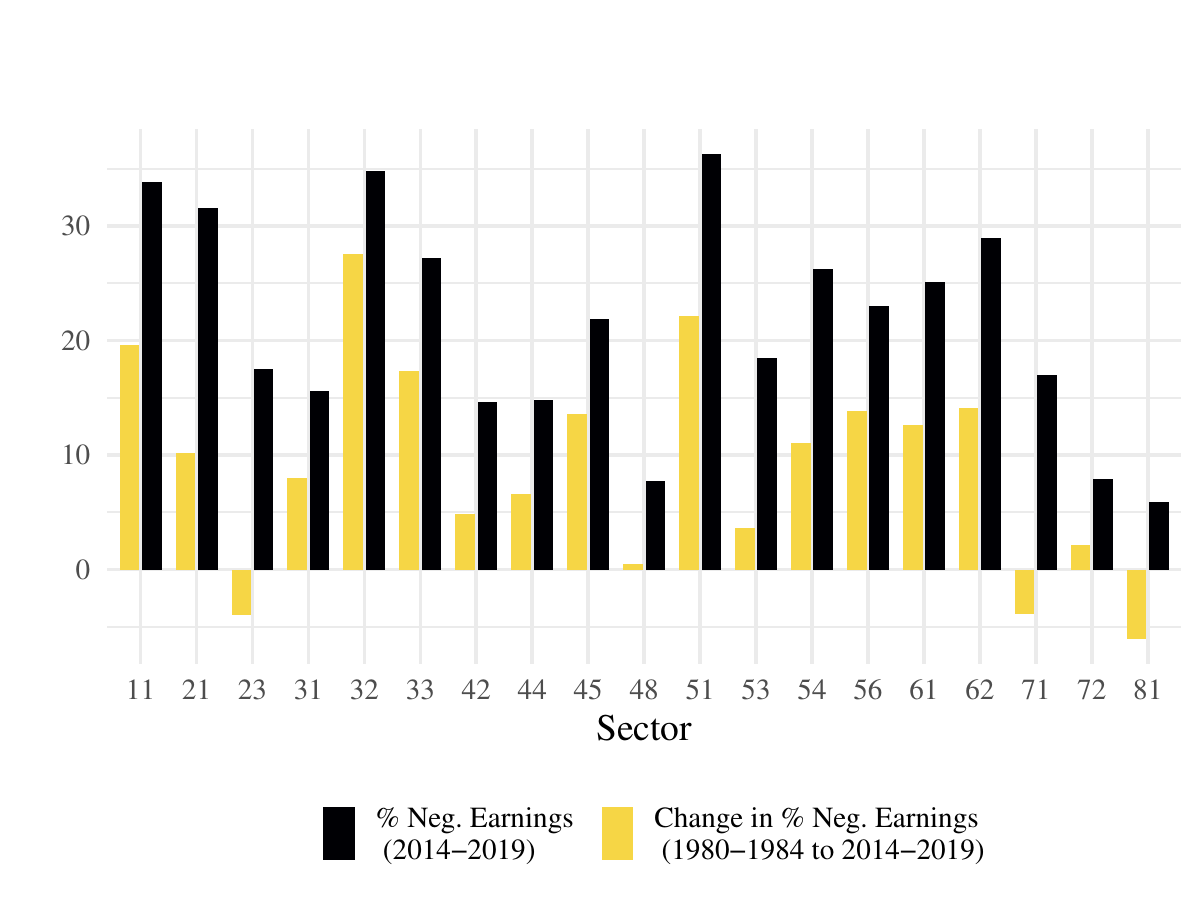}
    \caption{The Rise of Negative Earnings Across Sectors}
    \label{fig:neg_earnings_by_sector_2digit}
    \scriptsize\raggedright{\textit{The figure plots the percent of Compustat firms in our sample reporting losses averaged between 2014-2019, and the change relative to the average reporting losses between 1980-1984, for 2-digit NAICS industries.}}
\end{figure}

The rise of negative earnings is also common across sectors, as shown in Figure~\ref{fig:neg_earnings_by_sector_2digit}. All 2-digit NAICS industries see a rise in the percent of firms reporting $\text{EBITDA} < 0$ between 1980 and 2019, with the exception of NAICS 23 (Construction) and NAICS 71 (Arts, Entertainment, and Recreation).

\begin{figure}[!h]
    \centering
    \includegraphics[width=0.95\textwidth]{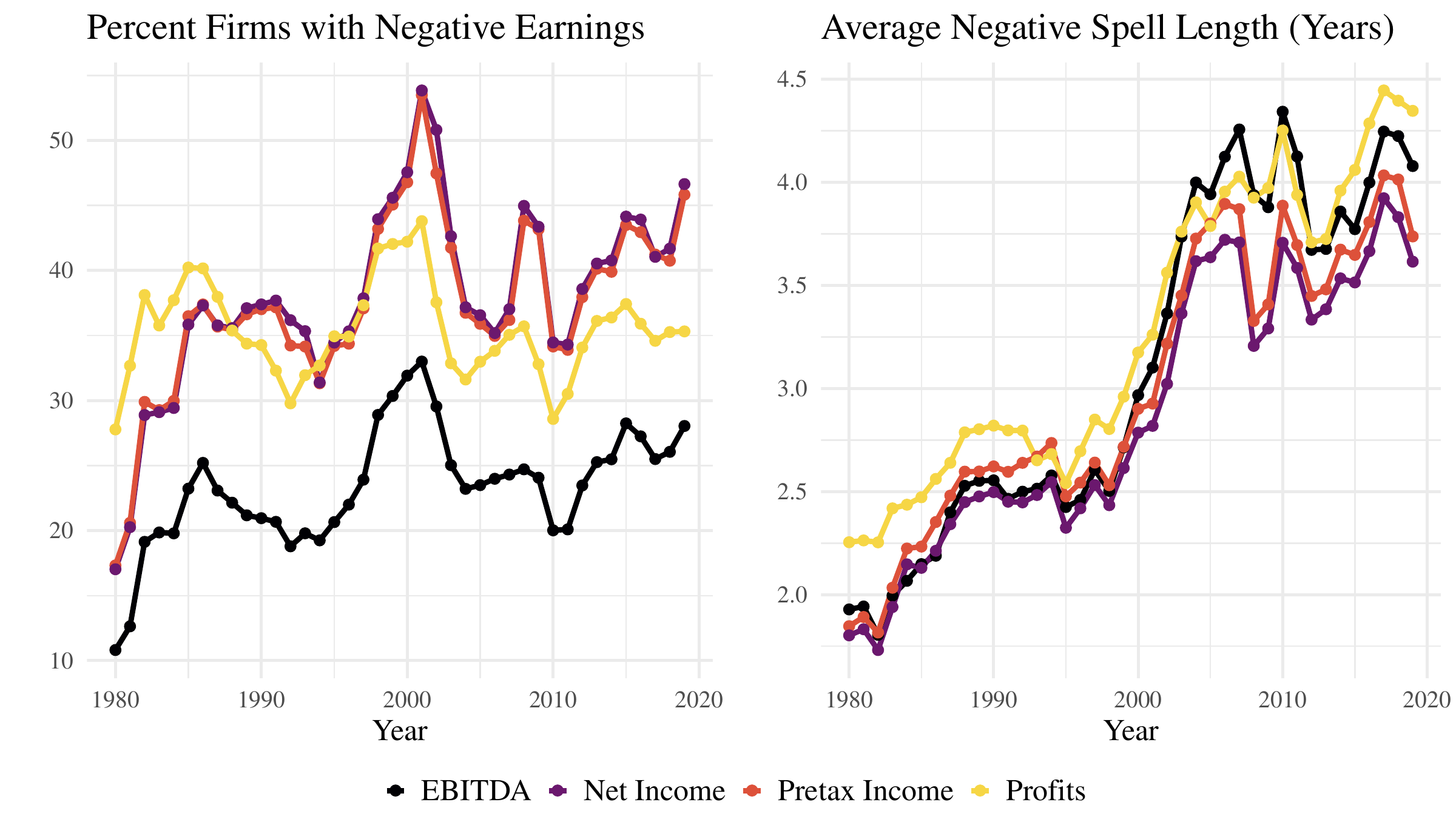}
    \caption{The Rise of Negative Earnings Across Earning Definitions}
    \label{fig:pct_negative_earnings_alt_measures}
     \scriptsize\raggedright{\textit{The figure plots the percent of Compustat firms in our sample reporting losses for each year (left panel) and the average number of consecutive years firms with losses have reported losses (right panel) between 1980 and 2019, for different definitions of earnings.}}
\end{figure}

\begin{figure}[!h]
    \centering
    \includegraphics[width=\textwidth]{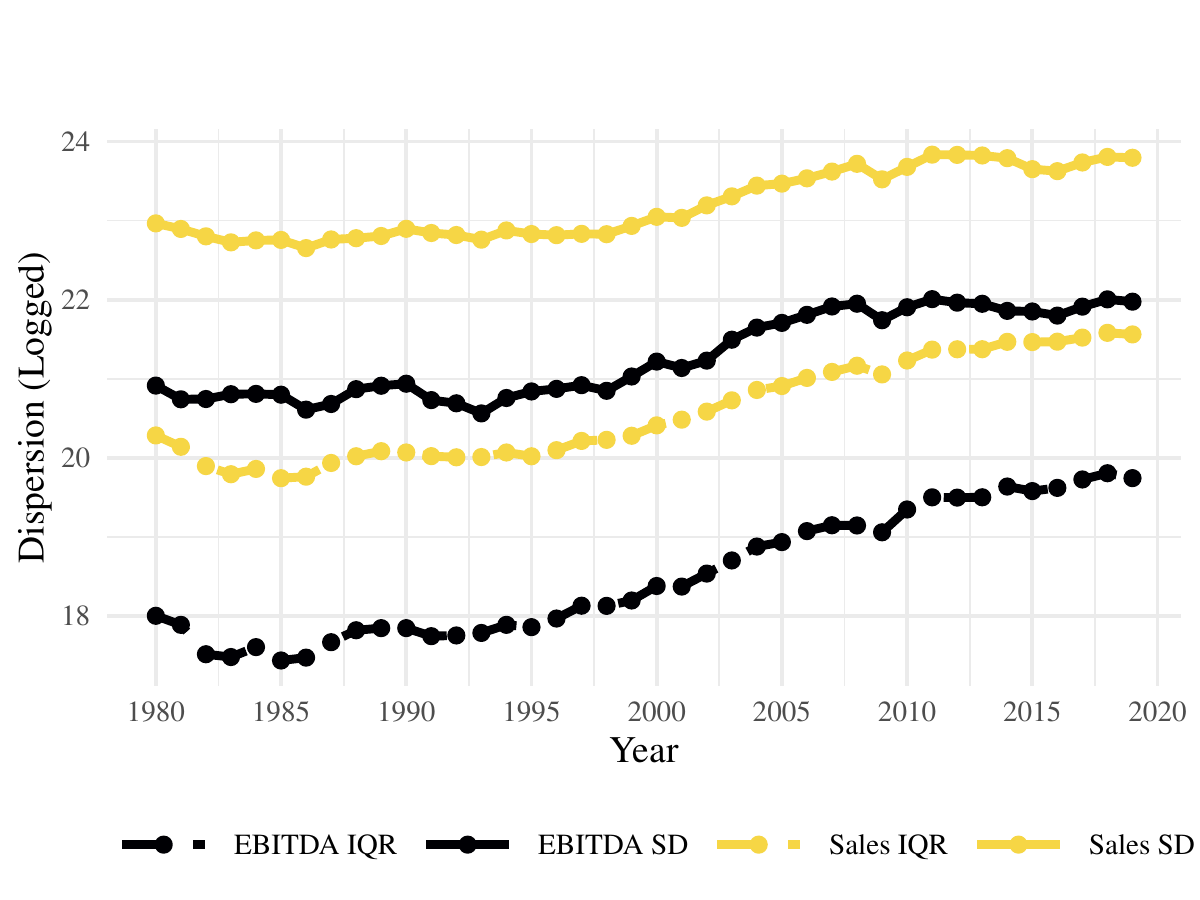}
    \caption{The Spreading of the Sales and Earning Distributions}
    \label{fig:sd_joint_by_year}
    \scriptsize\raggedright{\textit{The figure plots the standard deviation and inter-quartile range of earnings (EBITDA) and sales for Compustat firms in our sample, for each year between 1980-2019.}}
\end{figure}

The rise of negative earnings has occurred alongside a rise in real earnings: between 1980 and 2019, average and median EBITDA rose by a factor of \getstat{avg_earnings_factorchange} and \getstat{med_earnings_factorchange} respectively. These two trends have occurred simultaneously because the the sales and earnings distributions have spread: as shown in Figure~\ref{fig:sd_joint_by_year}, the standard deviations of real sales and real EBITDA rose by \getstat{sd_sales_percchange}\% and \getstat{sd_ebitda_percchange}\% respectively. This rise was remarkably uniform: as shown in Figures \ref{fig:top_qtiles_sales_earnings} and \ref{fig:bottom_qtiles_sales_earnings}, each quantile of the sales and EBITDA distribution pulled away from the median between 1980 and 2019.\footnote{Scale invariant measures of the earnings and sales distribution are shown in Figure \ref{fig:scale_inv_sales_earnings}.}

\paragraph{The Evolution of Firm Spending}

\begin{figure}[!h]
    \centering
    \includegraphics[width=\textwidth]{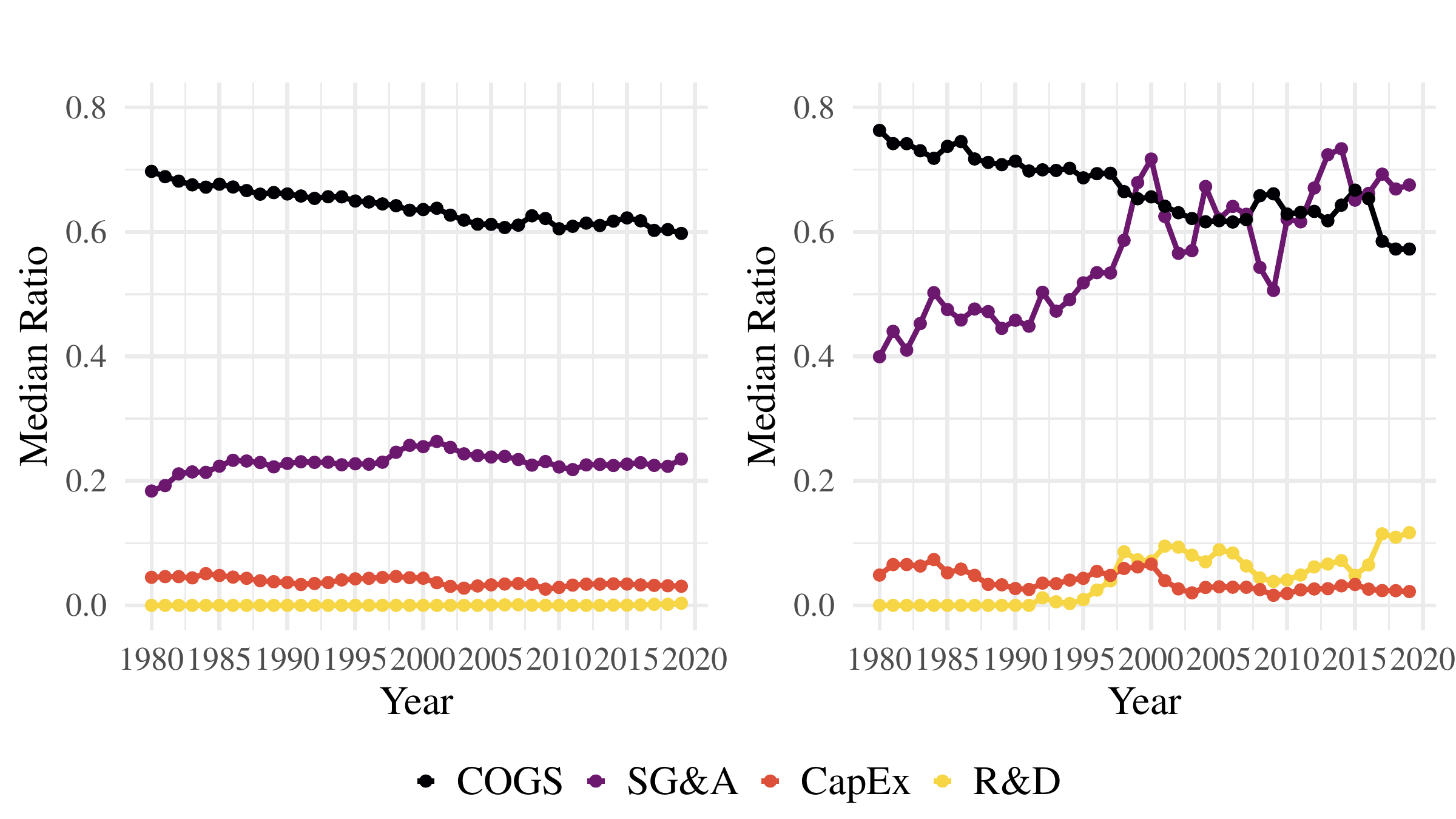}
    \caption{The Recomposition of Firm Spending: All Firms and Firms with Losses}
    \label{fig:cost_ratios_2panel}
    \scriptsize\raggedright{\textit{The figure plots spending on COGS, SG\&A, CapEx, and R\&D as a share of sales for Compustat firms in our sample (left panel) and for Compustat firms in our sample reporting losses that year (right panel), for each year between 1980-2019.}}
\end{figure}

Between 1980 and 2019 firm spending also transformed, as shown in Figure~\ref{fig:cost_ratios_2panel}. COGS and CapEx fell as a share of sales for the median firm by \getstat{cogs_change_all}\% and \getstat{capex_change_all}\% respectively. In contrast, SG\&A spending rose as a share of sales by \getstat{sga_change_all}\%, making it the only spending category to grow relative to revenue for the median firm.

This recomposition of spending is even more stark for firms with negative earnings. COGS and CapEx fell as a share of sales for the median firm with $\text{EBITDA} < 0$ by \getstat{cogs_change_neg}\% and \getstat{capex_change_neg}\% respectively. For the same firm, the SG\&A spending share rose by \getstat{sga_change_neg}\%, while the R\&D spending share rose by \getstat{rd_change_neg}\%.

In our model, we interpret this rise in SG\&A as a rise in sales and marketing spending. However, SG\&A is an extremely broad spending category, encompassing many other administrative costs. Using sector-level data from \citet{heInvestingCustomerCapital2024}, we show in Table \ref{tab:sector_regs} that the sales and marketing component of SG\&A is correlated across 3-digit NAICS industries with (1) the percent of firms reporting losses, (2) the persistence of losses, and (3) the spread of the EBITDA distribution. We see this as suggestive evidence that sales and marketing spending is what ties SG\&A to the rise of negative earnings. 

\paragraph{Summary}

In summary, between 1980 and 2019 negative earnings became a more common and more persistent state in our sample of public firms, with IRS SOI data implying the phenomena generalizes to the universe of US corporations. This occurred even as real earnings rose in our sample, because the distribution of sales and earnings spread evenly on either side of the median between 1980 and 2019: firms today make and lose more money than they did fifty years ago.

As earnings changed, so too did spending: COGS and CapEx fell as a share of sales for the median firm between 1980 and 2019, while the sales share of SG\&A spending rose. These trends were more pronounced for the median firm with negative EBITDA.

\section{Model}
\label{sec:theory}

In this section, we present a model of heterogenous firms engaging in supply and demand shifting investment that can rationalize the trends discussed in Section \ref{subsec:empirical_results}. Firms are heterogenous in their productivities a la \citet{melitzImpactTradeIntraIndustry2003} and \citet{hopenhaynJobTurnoverPolicy1993}. Supply shifting investment is modeled as a traditional physical capital stock, while demand shifting investment is modeled as endogenous customer acquisition a la \citet{afrouziConcentrationMarketPower2023}, extended to allow for scale effects in demand.

\subsection{Setup}

Time is discrete and indexed by $t$. The representative household has a continuum of members indexed by $j \in [0, 1]$. A continuum of monopolistically competitive firms indexed by $i \in [0,1]$ produce weakly substitutable goods.

\paragraph{Households}

The representative household supplies one unit of labor inelastically. Its members consume the different variety goods produced by the firms. Each period, each household member is a customer of a single firm $i$. $M_{it}$ represents the measure and the set of customers of firm $i$, i.e. $j \in M_{it}$ when household member $j$ is a customer of firm $i$ in period $t$, and $M_{it} = \int_0^1 \mathbf{1}_{j \in M_{it}} dj$.

Household members choose how much to consume by jointly maximizing the household consumption basket, a CES aggregator across varieties, each period subject to their budget constraint.
\begin{equation}
\label{eq:hh_problem}
\begin{aligned}
\max_{C_{ijt}} \space C_t \text{  s.t.  } \int_0^1 \int_0^1 P_{it} C_{ijt} dj\, di = W_t + \int_0^1 \tilde \Pi_{it} di \\
\text{Where } C_t = \left(\int_0^1 \int_0^1 \mathbf{1}_{j \in M_{it}}M_{it}^{\frac{\phi}{\sigma}}C_{ijt}^{\frac{\sigma-1}{\sigma}} dj\, di \right)^{\frac{\sigma}{\sigma - 1}}
\end{aligned}
\end{equation}
Here $C_t$ is the household consumption basket, $C_{ijt}$ is the consumption by household member $j$ of good $i$, $P_{it}$ is the price of good $i$, $W_t$ is the wage and $\tilde \Pi_{it}$ is the net profits of firm $i$, which include investment costs. Without loss of generality, normalize the price of the household consumption bundle $C_t$ to be $P_t = 1$. Because all household members have identical preferences, all customers of firm $i$ choose to purchase the same amount given price $P_{it}$.
\begin{equation*}
C_{ijt} = \mathbf{1}_{j \in M_{it}}M_{it}^{ \phi}C_tP_{it}^{-\sigma}
\end{equation*}
It's simple to recover total demand for good $i$, $C_{it}$, by integrating across customers.
\begin{equation*}
C_{it} = M_{it}^{1 + \phi} P_{it}^{-\sigma} C_t
\end{equation*}
Consumption decisions are shaped by two exogenous parameters: $\sigma$, the substitution elasticity, and $\phi$. By CES aggregation, $\sigma$ is the price elasticity of demand. Similarly, $\phi$ is the customer base elasticity of demand: $\frac{\partial \log C_{ijt}}{\partial \log M_{it}} = \phi$, for $j \in M_{it}$. We include this parameter to allow demand per customer to depend on number of customers, and we refer to it as the \emph{scale elasticity of demand}, since it generates scale effects in customer accumulation.

This does imply that the household consumption basket depends on consumption quantities $C_{ijt}$ mediated by the size of firms' customer bases $M_{it}$ according to the scale elasticity of demand $\phi$. To ground our intuition, consider social media companies like Meta. If we assume $\phi > 0$, our model implies that the household gets more out of consuming Instagram when Instagram has more users.

\paragraph{Customer Base}

Each period $t$, $\theta$ firms exit randomly and are replaced by $\theta$ new firms, which inherit $\underline{m}$ customers. In addition, $\delta_m$ customers at each surviving firm separate. This process leaves free customers of measure $\theta (1 - \underline{m}) + (1-\theta) \delta_m$.

Free customers are acquired by advertisements that firms post in the previous period, which we denote as $a_{it-1}$. Firm $i$'s customer base evolves according to
\begin{equation*}
M_{it} = (1-\delta_m)M_{it-1} + \frac{a_{it-1}}{P_{mt}}
\end{equation*}
$P_{mt}$ is the endogenous conversion rate of advertisements to customers. It is determined in equilibrium to clear the free customer market.
\begin{equation*}
\begin{aligned}
P_{mt} = \frac{1-\theta}{\theta (1-\underline{m}) + (1-\theta) \delta_m} \int_0^1 a_{it-1}\, di
\end{aligned}
\end{equation*}

\paragraph{Firms}

At each period $t$, firms draw their productivity $Z_{it}$. Entrant firms draw initial productivity from a log-normal distribution
\begin{equation*}
\ln Z_{it} \sim \mathbf{N} (\bar Z, \sigma_Z^2)
\end{equation*}
Incumbent firms draw productivities according to an AR(1) process
\begin{equation*}
\ln Z_{it} = \rho \ln Z_{it-1} + \epsilon_{it}, \quad \epsilon_{it} \sim \mathbf{N}(0, \sigma_Z^2)
\end{equation*}
After drawing their productivities, firms face a static profit maximization problem and a dynamic investment problem, both of which are defined by three state variables: customer base $M_{it}$, physical capital stock $K_{it}$, and productivity $Z_{it}$.

The static problem is standard monopolist price/quantity setting: given production cost function $C(Y_{it}, K_{it}, Z_{it})$,  firms solve
\begin{equation}
\label{eq:firm_problem_static}
\Pi_{it} (M_{it}, K_{it}, Z_{it}) = \max_{Y_{it}} P_{it} Y_{it} - C(Y_{it}, K_{it}, Z_{it}) \quad \text{where} \quad Y_{it} = M_{it}^{1 + \phi} P_{it}^{-\sigma} C_t
\end{equation}
Where the demand equation comes from the household problem described above, and firms take aggregate consumption $C_t$ as given. We assume Cobb-Douglas production with labor and capital multipliers $\gamma_l$ and $\gamma_k$.

Firms customer base $M_{it}$ evolves as described above, and their physical capital stock $K_{it}$ evolves according to the standard formula.
\begin{equation*}
K_{it} = (1-\delta_k) K_{it-1} + i_{it-1}
\end{equation*}
Where $i_{it-1}$ is physical capital investment. Firms produce advertisements $a_{it}$ and investment $i_{it}$ using labor, with decreasing returns to scale production. 
\begin{equation*}
a_{it}(L_{a,it}) = L_{a,it}^{\alpha_a}; \quad i_{it}(L_{k,it}) = L_{k,it}^{\alpha_k}
\end{equation*}

The capital and advertisement production functions can be rewritten as convex cost functions which depend on wages $W_t$. We consider spending on labor for capital production ($W_t L_{k,it}$) as capital expenditures, and spending on labor for advertisement production ($W_t L_{a, it}$) as sales and marketing spending. 

Thus, the dynamic investment problem of firm $i$ is represented by the following Bellman equation.
\begin{equation}
\label{eq:firm_problem_dynamic}
\small
\begin{aligned}
V(M_{it}, K_{it}, Z_{it}) = \max_{a_{it}, i_{it}} \pi(M_{it}, K_{it}, Z_{it}) - W_t\!\left(i_{it}^{\frac{1}{\alpha^k}} + a_{it}^{\frac{1}{\alpha_a}}\right) + \tilde \beta \mathbb{E}_t[V(M_{it+1}, K_{it+1}, Z_{it+1})] 
\\
\text{subject to } K_{it+1} = (1-\delta_k) K_{it} + i_{it}; \space \space M_{it+1} = (1-\delta_m) M_{it} + \frac{a_{it}}{P_{mt+1}}
\end{aligned}
\end{equation}
Where $\tilde \beta = \beta (1-\theta)$, i.e. discounting accounting for the survival probability.

\paragraph{Equilibrium}

We are interested in the steady state equilibrium of this model, the stationary distribution where agents optimize, aggregate values are fixed, and markets clear. More precisely, this equilibrium is defined as prices $\left[P_{it}, W_t\right]$ and quantities $\left[Y_{it}, C_{ijt}, L_{it}, L_{a,it}, L_{k,it}\right]$ such that

\begin{enumerate}
    \item Quantities $C_{ijt}$ satisfy the household problem in Equation~\ref{eq:hh_problem}, given prices.
    \item Quantities $Y_{it}, L_{it}$ satisfy the static firm problem in Equation~\ref{eq:firm_problem_static}, given prices.
    \item Quantities $L_{a,it}, L_{k,it}$ satisfy the dynamic firm problem in Equation~\ref{eq:firm_problem_dynamic}, given prices.
    \item Labor, quantity, and customer markets clear, and aggregates are constant (i.e. $C_{t+1} = C_t = C$, $P_{mt+1} = P_{mt} = P_m$).
\end{enumerate}

\subsection{Supply vs.\ Demand Shifting Investment}

A novel feature of our model is the ability of firms to invest in both traditional physical capital $K_{it}$ and their customer base $M_{it}$. In this section we discuss how these types of investment differ, which will be central in the quantitative exercises that follow. 

Returning to Equation~\ref{eq:firm_problem_dynamic}, the returns on an additional unit of physical capital $K_{it}$ and an additional customer $M_{it}$ are defined by the respective envelope conditions.
\begin{align*}
V_M(M_{it}, K_{it}, Z_{it}) &= \Pi_M(M_{it}, K_{it}, Z_{it}) + \tilde \beta (1-\delta_m) \mathbb E_t V_M(M_{it+1}, K_{it+1}; Z_{it+1}) \\
V_K(M_{it}, K_{it}; Z_{it}) &= \Pi_K(M_{it}, K_{it}, Z_{it}) + \tilde \beta (1-\delta_k) \mathbb E_t V_K(M_{it+1}, K_{it+1}; Z_{it+1})
\end{align*}
Each return thus depends recursively on the partial derivative of static profits with respect to customers and capital. Static profits can be written as a function of customers, capital, and productivity as follows.
\begin{equation*}
\Pi(M_{it}, K_{it}, Z_{it}) = \left(W_t^{\frac{1}{\Lambda} - 1} (Z_{it} K_{it}^{\gamma_k})^{\frac{1}{\gamma_l}\left(1 - \frac{1}{\Lambda}\right)} (M_{it}^{1+\phi} C_t)^{\frac{1}{\sigma}} \frac{\gamma_l \sigma}{\sigma-1} \right)^\Lambda\!\left(\frac{\sigma}{(\sigma-1)\gamma_l} - 1 \right)
\end{equation*}
Where $C_t$ is aggregate consumption, $W_t$ is the wage, and $\Lambda = \frac{\sigma}{(1-\gamma_l) \sigma + \gamma_l}$. The relevant partial derivatives can be expressed in terms of the corresponding elasticities.
\begin{equation*}
\Pi_M(M_{it}, K_{it}, Z_{it}) = \frac{\Pi(M_{it}, K_{it}, Z_{it})}{M_{it}} \epsilon^\Pi_M ; \quad \Pi_K(M_{it}, K_{it}, Z_{it}) = \frac{\Pi(M_{it}, K_{it}, Z_{it})}{K_{it}} \epsilon^\Pi_K
\end{equation*}
Where $\epsilon^\Pi_M$ and $\epsilon^\Pi_K$ are the elasticities of profits with respect to customers and capital correspondingly.
\begin{equation*}
\epsilon^\Pi_K =\frac{\gamma_k(\sigma - 1 )}{(1-\gamma_l) \sigma + \gamma_l}; \quad
\epsilon^\Pi_M = \frac {1+\phi} {(1-\gamma_l) \sigma + \gamma_l}
\end{equation*}
Thus the ratio of marginal profits, which determines the ratio of marginal returns and the ratio of investment intensities, has a simple analytic expression.
\begin{equation}
\label{eq:relative_returns}
\frac{\Pi_M(M_{it}, K_{it}, Z_{it})}{\Pi_K(M_{it}, K_{it}, Z_{it})} = \frac{K_{it}}{M_{it}} \times \frac{1+\phi}{\gamma_k (\sigma-1)}
\end{equation}
With these results in mind, we can characterize the difference between physical investment $i_{it}$ and customer investment $a_{it}$ as threefold.

\begin{figure}[!h]
    \centering
    \includegraphics[width=.75\textwidth]{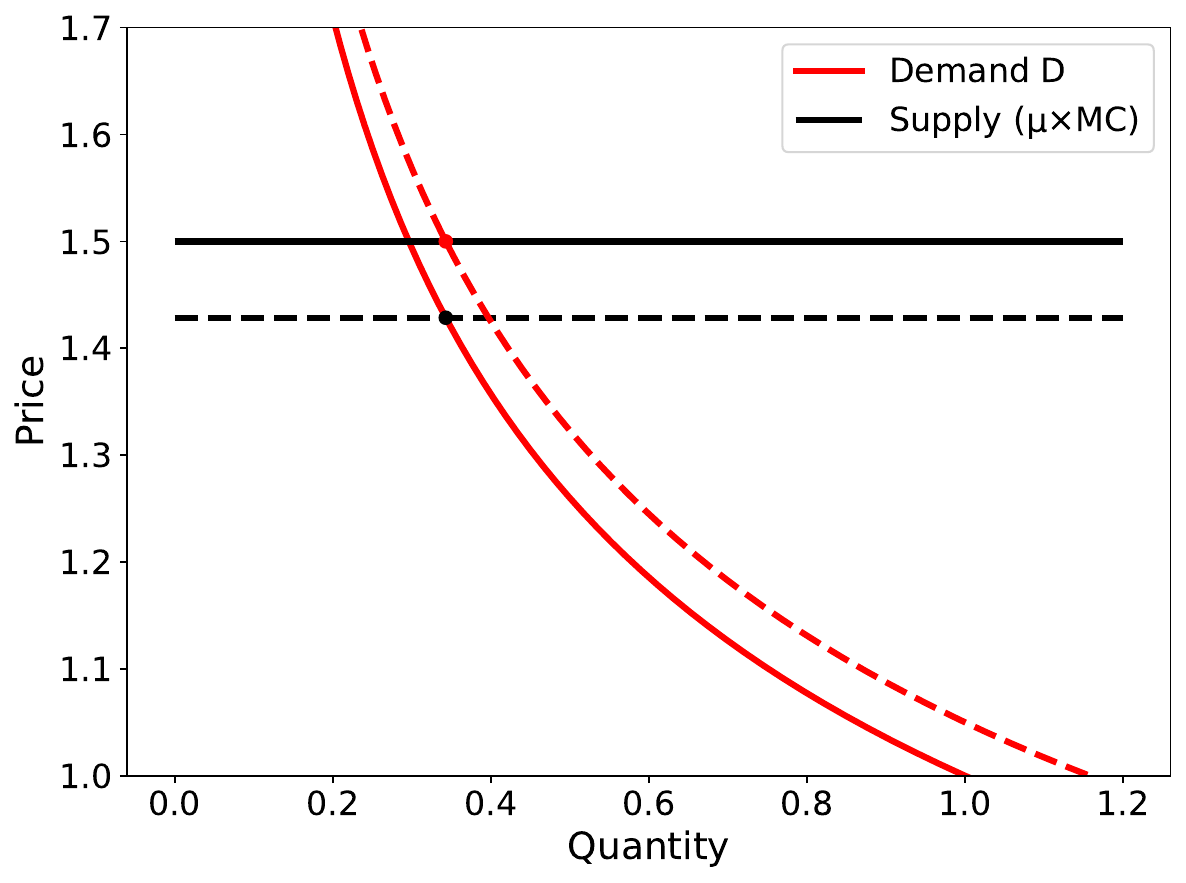}
    \caption{Shifting Supply Curves vs. Shifting Demand Curves}
    \label{fig:stylizedSEP}
\end{figure}

First, \emph{customer investment shifts the demand curve, while physical investment shifts the supply curve}, as shown in Figure~\ref{fig:stylizedSEP}. Put differently, customer investment directly affects quantity demanded. Advertisements $a_{it}$ translate to customers in the next period, which directly scale the quantity demanded at any given price. Physical investment indirectly affects quantity demanded through price. Investment $i_{it}$ lowers the marginal cost of production in the next period, lowering the firms optimal price choice and raising quantity demanded according to the household's demand curve.

Second, as a consequence of this the return on physical investment $i_{it}$ depends directly on the \emph{price elasticity of demand}, while the return on customer investment $a_{it}$ depends directly on the \emph{scale elasticity of demand}.\footnote{Both returns depend indirectly on the price elasticity when static production isn't CRS, i.e. $\gamma_l \neq 1$.} This follows from Equation~\ref{eq:relative_returns}, and the intuition can be seen in Figure~\ref{fig:stylizedSEP}. Physical investment increases quantity demanded by lowering prices. The returns on this investment depend on the relationship between price and quantity demanded, i.e. the price elasticity of demand $\sigma$. Customer investment increases quantity demanded by increasing the number of customers. The returns on this investment depend on the relationship between demand per customer and number of customers, i.e. the scale elasticity of demand $\phi$.

Third, \emph{customer investment is pure business stealing}, while physical investment is not. More precisely, all equilibrium quantities except $a_{it}$ are homogenous of degree 0 in aggregate advertising. This result follows from customer market clearing, which determines the endogenous conversion rate of advertisements to new customers. Since $P_{mt} = \kappa \int_0^1 a_{it-1}\, di$ as derived above, scaling advertising scales the conversion rate, leaving all other equilibrium quantities unchanged.

Intuitively, customer investment determines \emph{relative demand}, not aggregate demand. It can (and does) affect aggregate output, but only by determining how demand is allocated across firms: its aggregate level is irrelevant. In contrast, physical investment increases output irrespective of the actions of other firms. Its aggregate level affects aggregate output by determining the production possibilities frontier.

\section{Quantitative Results}
\label{sec:quantitative}

In this section we solve for a steady state equilibrium of our model corresponding to each year in our sample. We are able to quantitatively match the rise in the percent of firms with negative earnings and qualitatively match the other trends discussed in Section \ref{subsec:empirical_results} using the scale elasticity of demand $\phi$ as the only driver of changes across equilibria. We then construct a measure of customer capital $M_{it}$ in our data using the perpetual inventory method, and use this measure to present evidence that the scale elasticity of demand has indeed risen over our sample period, in line with its calibrated path.

\subsection{Calibration and Solution Strategy}

\paragraph{Calibration}

Our model has 11 exogenous parameters, summarized in Table~\ref{tab:calibration}. We leave the scale elasticity of demand $\phi$ as a free parameter: we set it to 0 in our 1980 equilibrium, and then calibrate it to match the percent of firms reporting negative earnings in each year that follows.

\begin{table}[!h]
    \centering
    \caption{Calibration Table}
    \label{tab:calibration}
\begin{tabular}{lc}
\toprule
\textbf{Parameter} & \textbf{Value} \\
\midrule
Discount rate $\beta$ & 0.96 \\
Persistence of productivity $\rho_z$ & 0.82 \\
SD of productivity $\sigma_z$ & 0.04 \\
Entry/Exit rate $\theta$ & 0.11 \\
Depreciation rates $\delta_k, \delta_d$ & 0.10, 0.15 \\
Output elasticities $\gamma_L, \gamma_K$ & 0.86, 0.11 \\
Price elasticity $\sigma$ & 3.15 \\
Investment RTS $\alpha_k, \alpha_a$ & 0.83, 0.46 \\
\bottomrule
\end{tabular}
\end{table}

We directly observe the entry/exit rate $\theta$ in our Compustat sample. We use a weighted average of our sectoral production functions, estimated using the ACF methodology, to calibrate the Cobb-Douglas parameters $\gamma_k$ and $\gamma_l$ along with the persistence and variance of the productivity process $\rho$ and $\sigma_z$. We also estimate the median mark-up $\mu$ with the ratio estimator, and use it to back out the elasticity of substitution $\sigma$.

We set the discount factor $\beta = 0.96$, the standard value for annual data. We set $\delta_k = 0.1$, and $\delta_m = 0.15$ in accordance with \citet{gourioCanIntangibleCapital2014} and \citet{Dhyne2023Belgian}. Finally, we calibrate the returns to scale of advertisement and investment production $\alpha_a$ and $\alpha_k$ to match (1) the percent of firms reporting negative earnings in 1980 and (2) the median ratio of capital expenditures over sales in 1980.

\paragraph{Solution}

We solve for the steady state equilibrium by (1) solving the firm problem given steady state aggregate consumption $C$, wage $W$, and conversion rate $P_{m}$, and (2) globally searching for the fixed point $X^* = (C^*, W^*, P_m^*)$ where $X(g(X^*)) = X^*$, i.e. the values used to solve the firm problem clear the product, labor, and customer markets in the stationary distribution.

The core computational challenge is solving for the firm policy function given a guess of steady state values efficiently enough to globally search for this fixed point. To accomplish this, we use the endogenous grid method of \citet{carrollMethodEndogenousGridpoints2006}. We discretize our state variable spaces, and take an initial guess at the value function derivatives $V_k$ and $V_a$. Given these guesses, we backwards iterate to the endogenous grid of state variables that correspond to our discretized grid using the firm first order conditions. We interpolate to recover policy choices on our discretized grid, which we use to generate updated guesses $V_k', V_a'$. More details on our computational strategy can be found in the Appendix.

\subsection{Quantitative Results}

\begin{figure}[!h]
    \centering
    \includegraphics[width=\textwidth]{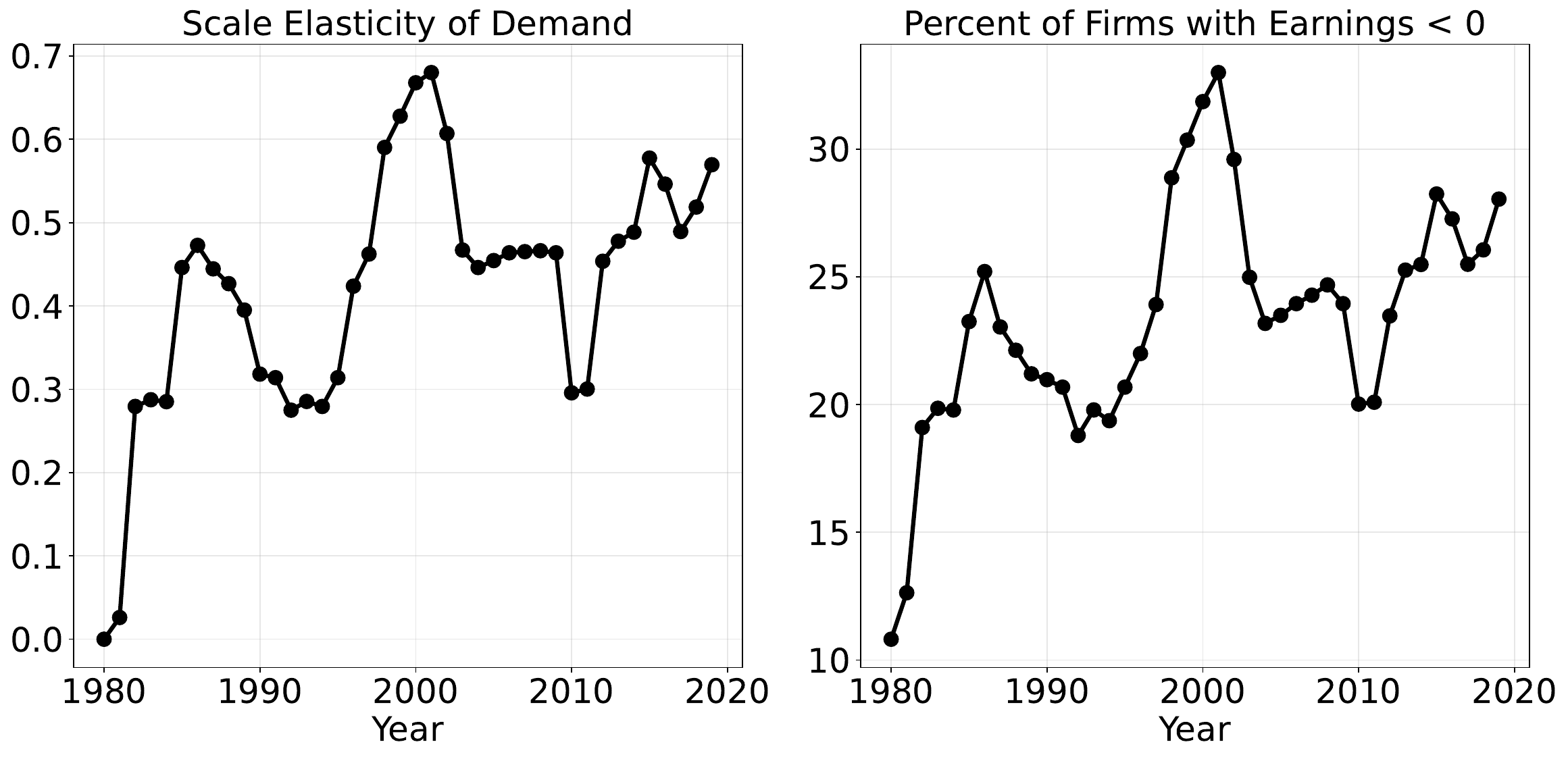}
    \caption{The Scale Elasticity and the \% of Firms with Negative Earnings}
    \label{fig:phi_and_pct_neg_by_year}
    \scriptsize\raggedright{\textit{The figure plots a scale elasticity (left panel) and the \% of firms with negative earnings in the corresponding steady state equilibrium (right panel). The scale elasticity is calibrated to match the model moment with moment observed in our Compustat sample for each year 1980-2019.}}
\end{figure}

We solve for steady state equilibria corresponding to each year in our data. We use the scale elasticity of demand $\phi$ as the single driver of changes across equilibria, setting it to 0 initially and moving it to match the percent of firms with negative earnings each year between 1980 and 2019.\footnote{Earnings in our model are defined as $P_i Y_i - W (L_i + L_{a, i})$. We exclude expenses on labor used for physical investment, in line with the definition of EBITDA. We plot the implied time series for profits, $P_i Y_i - W(L_i + L_{a, i} + L_{k, i})$ in Figure \ref{fig:neg_ebitda_neg_profits}.} As Figure~\ref{fig:phi_and_pct_neg_by_year} shows, we can perfectly match this moment in the data with changes in the scale elasticity alone.

This is for two reasons. First, raising the scale elasticity raises firms' perceived return on investing in customers relative to investing in physical capital, changing investment intensities accordingly. Since customer investment is business stealing, firms over-invest and a larger portion end up losing money. Second, raising the scale elasticity increases the optimal customer base size. Firms want to grow larger over their lifetime which, given the convexity of investment cost due to the concavity of investment production functions, pushes them to spend more on forward looking investment for longer. 

\begin{figure}[!h]
    \centering
    \includegraphics[width=\textwidth]{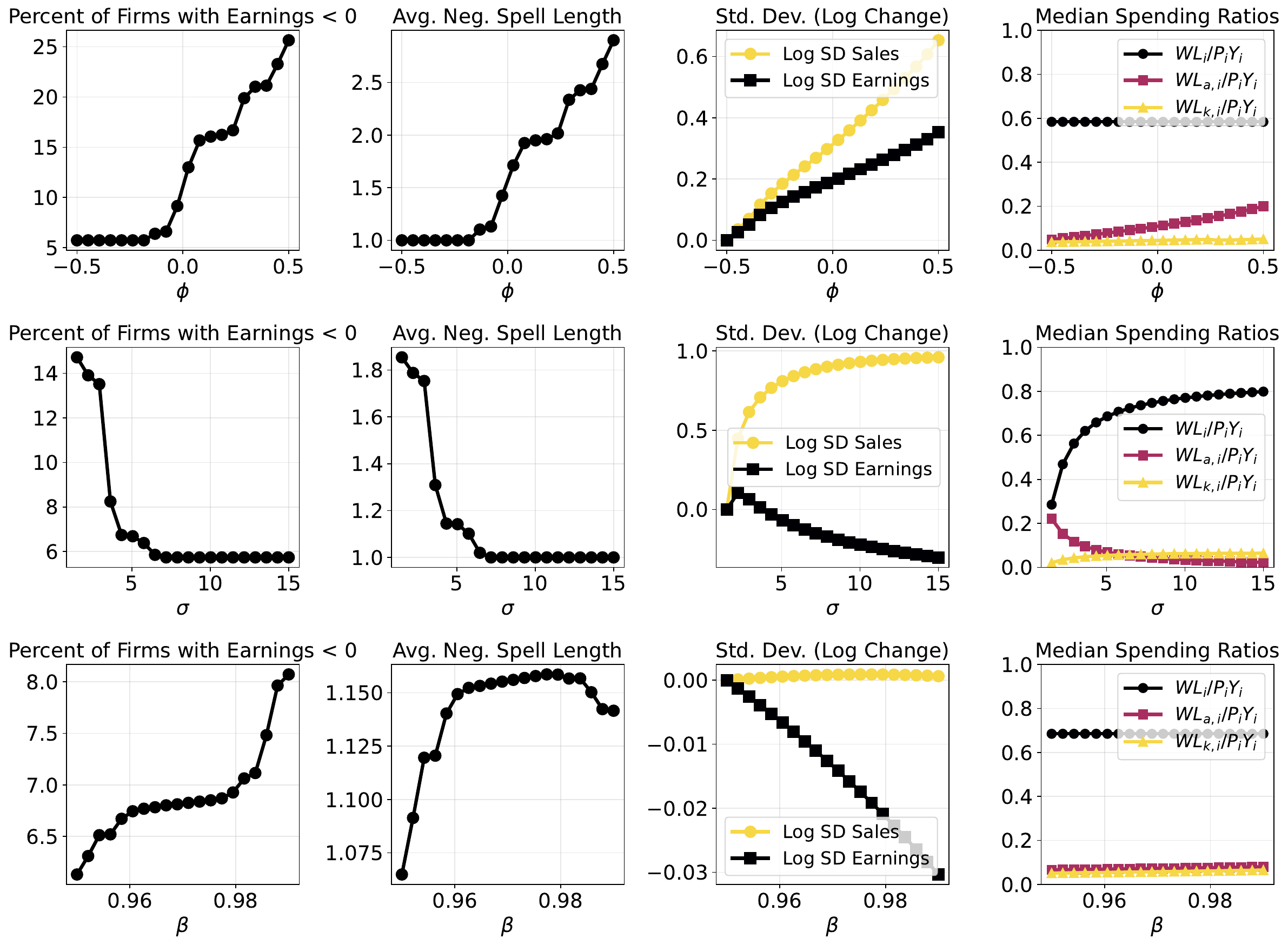}
    \caption{Model Moments vs. Parameters}
    \label{fig:moments_v_params}
    \scriptsize\raggedright{\textit{The figure plot four moments of the steady state equilibrium corresponding to our empirical trends discussed in Section \ref{subsec:empirical_results}, varying the scale elasticity (row 1) substitution elasticity (row 2), and discount rate (row 3).}}
\end{figure}

These mechanisms mean that the percent of firms with negative earnings is monotonically increasing in the scale elasticity of demand across equilibria, \emph{along with three other moments of the stationary distribution of firms}: (1) the average negative earning spell, (2) the sales and earnings distribution spread, and (3) sales and marketing spending relative to other costs. As shown in Figure \ref{fig:moments_v_params}, this is unique to the scale elasticity relative to other parameters, e.g. the elasticity of substitution $\sigma$ and the discount rate $\beta$. As a consequence, we are able to both quantitatively match the rise in the percent of firms with negative earnings and qualitatively match the other trends described in Section \ref{subsec:empirical_results} using changes in one parameter: the scale elasticity of demand.

\begin{figure}[!h]
    \centering
    \includegraphics[height=.9\textheight]{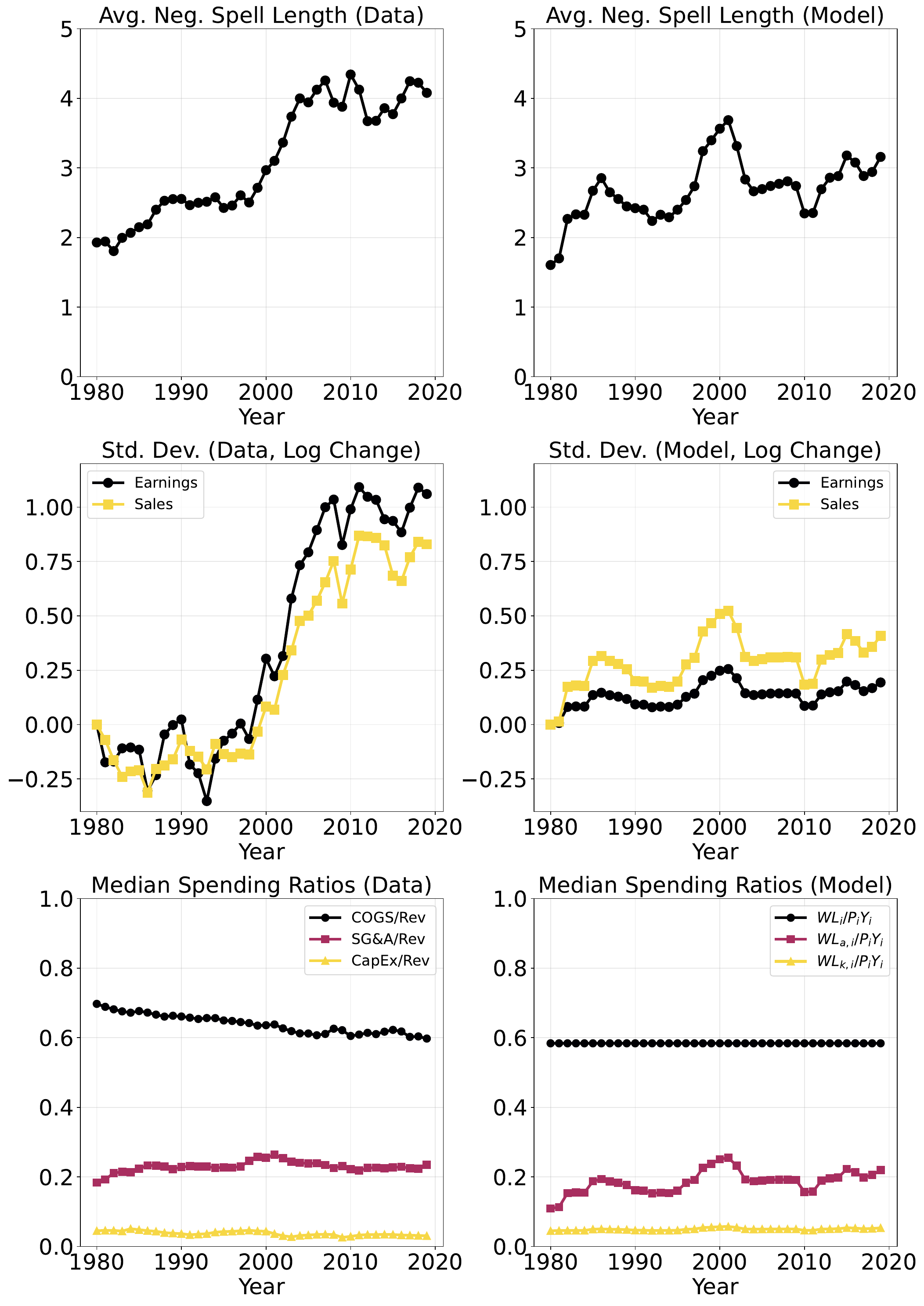}
    \caption{Untargeted Moments, Data vs. Model}
    \label{fig:untargeted_moments}
    \scriptsize\raggedright{\textit{The figure plots the three untargeted moments of the data, as observed in our Compustat sample (left panels) and in the steady state equilibria of our model corresponding to the calibrated path of the scale elasticity (right panels).}}
\end{figure}

The intuition for this result is as follows: First, since raising the scale elasticity pushes firms to invest more for longer, it raises the average negative earning spell, the number of consecutive periods a firm with negative earnings has had negative earnings. As shown in the first row of Figure~\ref{fig:untargeted_moments}, this rise corresponds to the rise observed in the data: in the model the average negative earning spell rises from \getstat{neg_spell_1980_model} to \getstat{neg_spell_2019_model}, while in the data the average negative earning spell rises from \getstat{neg_spell_1980} to \getstat{neg_spell_2019}.

Second, since raising the scale elasticity increases the optimal lifetime size of firms, it stretches out the sales and earnings distributions. The implied increases in the standard deviation of both sales and earnings qualitatively align with the increase we see in the data, though the magnitudes differ meaningfully. As shown in the second row of Figure~\ref{fig:untargeted_moments}, in the model the standard deviation of earnings (sales) rises by \getstat{sd_earnings_percchange_model}\% (\getstat{sd_sales_percchange_model}\%) between 1980 and 2019, while in the data it rises by \getstat{sd_ebitda_percchange}\% (\getstat{sd_sales_percchange}\%). 

Third, since raising the scale elasticity increases the return to investing in customers relative to investing in physical capital, it recomposes spending accordingly. The increase in the median spending rate on advertising, relative to production costs and capital investment, mirrors the the recomposition of spending we see in the data. As shown in third row of Figure~\ref{fig:untargeted_moments}, in the model advertising spending as a share of sales rises from \getstat{adv_ratio_1980_model} to \getstat{adv_ratio_2019_model}, while production costs as a share of sales remains flat and investment as a share of sales rises by \getstat{inv_ratio_change_model}. In the data, SG\&A spending as a share of sales rises by \getstat{sga_change_all}, while COGS and CapEx fell as a share of sales for the median firm by \getstat{cogs_change_all} and \getstat{capex_change_all} respectively.

\subsection{Has the Scale Elasticity of Demand Risen?}

Given that changes in the scale elasticity of demand $\phi$ drive all the results listed above, it is natural to ask whether there is any evidence that $\phi$ has actually changed between 1980 and 2019. To answer this question, we attempt to estimate the revenue elasticity of customer capital $M_{it}$, which is the same as the profit elasticity described in Section 2.2, and is therefore linear in $\phi$. In particular, our model implies the following log-linear relationship.
\begin{equation*}
\log P_{it} Y_{it} = \kappa + \epsilon^\Pi_M \log M_{it} + \epsilon^\Pi_K \log K_{it} + \xi_{it}
\end{equation*}
Where $\xi_{it} \propto \log Z_{it}$, $\kappa$ is a constant, and most importantly, $\epsilon^\Pi_M \propto \phi$. Since the revenue elasticity of customer capital is proportional to the scale elasticity of demand, we can test for changes in the scale elasticity by estimating this revenue elasticity for each year in our sample. However, to estimate this relationship we need to a measure of customer capital $M_{it}$.

We construct this measure using the perpetual inventory method \citep{eisfeldtOrganizationCapitalCrossSection2013, petersIntangibleCapitalInvestmentq2017}. Consistent with our model, we estimate customer capital $M_{it}$ by accumulating past SG\&A spending, \textbf{normalized by aggregate SG\&A spending within the market-year}. As noted above, we separate R\&D spending from SG\&A as in \citet{petersIntangibleCapitalInvestmentq2017}. We use Compustat data on our firm sample stretching back to 1950.
\begin{equation*}
M_{it} = (1-\delta_M) M_{it-1} + \frac{a_{it-1}}{P_{mt}}; \quad a_{it-1} \propto \text{SG\&A}_{it}, \space P_{mt} \propto \int_0^1 a_{it-1}\, di
\end{equation*}
A challenge in applying the perpetual inventory method is picking $M_{i0}$, the customer capital stock corresponding to the first time firm $i$ appears in the Compustat record. We follow the methodology of \citet{petersIntangibleCapitalInvestmentq2017} to impute this value, using data on firms' founding years from Jay Ritter's website, and data on firms' pre-IPO investment growth rates from Compustat. Further details are available in the Appendix.

With an empirical proxy for customer capital $M_{it}$ in hand, we test whether the revenue elasticity of customer capital has evolved over our sample period using the a regression specification corresponding to the model-implied equation above.
\begin{equation}
\label{eq:sales_elasticity_reg}
\log \text{Sales}_{it} = \beta_t \log M_{it} + \alpha \log K_{it} +\gamma_{st} + \gamma_i
\end{equation}
I.e., we regress logged sales on logged customer capital interacted with year, logged physical capital proxied using Compustat variable ppegt, and sector-year and firm fixed effects. $\beta_t$ is our estimate of $\epsilon^\Pi_t$, the revenue elasticity of customer capital in year $t$. Assuming all other exogenous parameters are fixed, the evolution of this estimate between 1980 and 2019 is proportional to the evolution of the scale elasticity of demand.

\begin{figure}[!h]
    \centering
    \includegraphics[width=\textwidth]{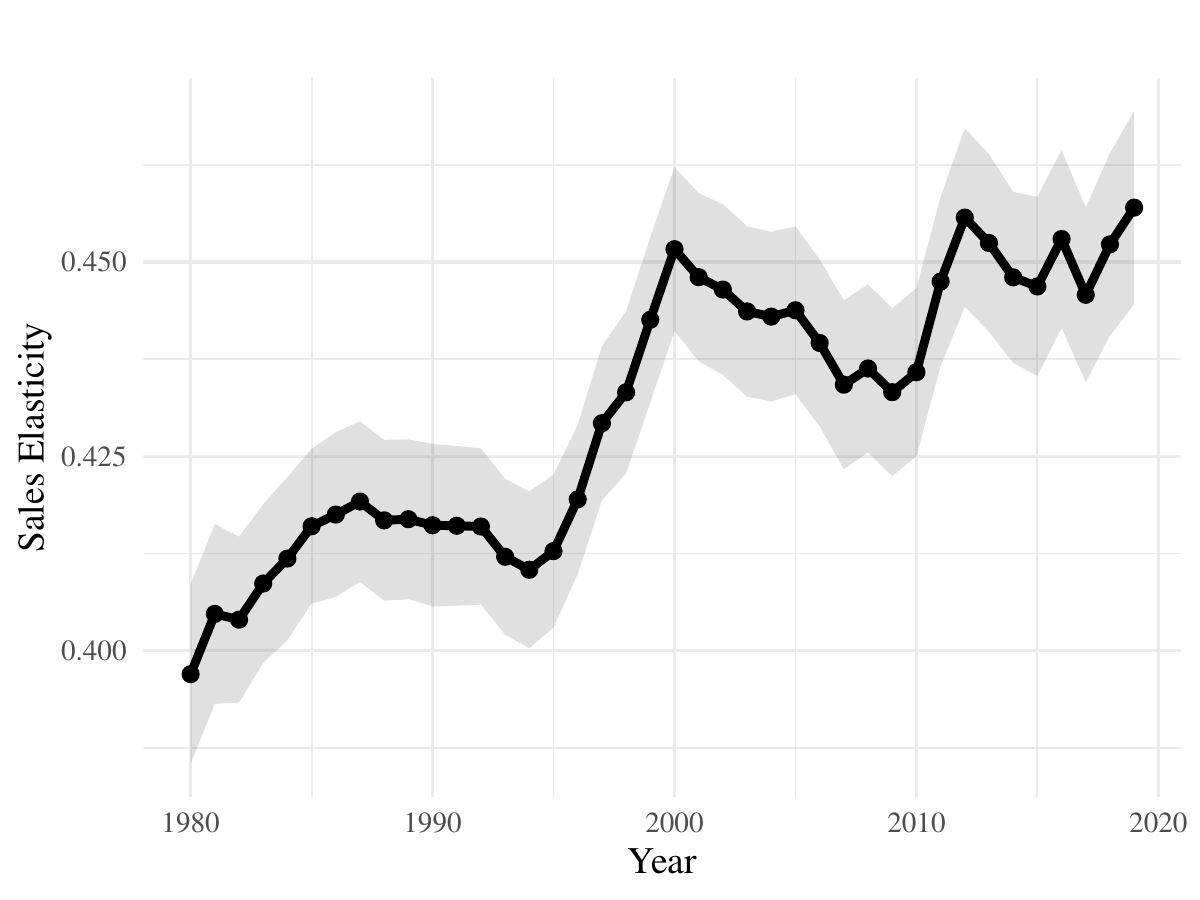}
    \caption{The Sales Elasticity of Customer Capital}
    \label{fig:sales_elasticity_m_by_year}
    \scriptsize\raggedright{\textit{The figure plots the estimated sales elasticity of customer capital for our sample of Compustat firms, for each year between 1980-2019. Our estimates are the year-specific coefficients on our proxy of customer capital, as described in the regression specification in Equation \ref{eq:sales_elasticity_reg}.}}
\end{figure}

As shown in Figure~\ref{fig:sales_elasticity_m_by_year}, this estimate implies that the scale elasticity of demand did indeed rise between 1980 and 2019. Not only that, its evolution closely mirrors the rise in percent of firms reporting negative earnings, and therefore the model-implied path of the scale elasticity of demand shown in Figure~\ref{fig:phi_and_pct_neg_by_year}. We see this as strong suggestive evidence of our model's mechanism.

\section{Aggregate Implications}
\label{sec:aggregation}

In this section, we consider the effect of raising the scale elasticity of demand on aggregate consumption, investment, and advertising. 

\begin{figure}[!h]
\centering 
\includegraphics[width=\textwidth]{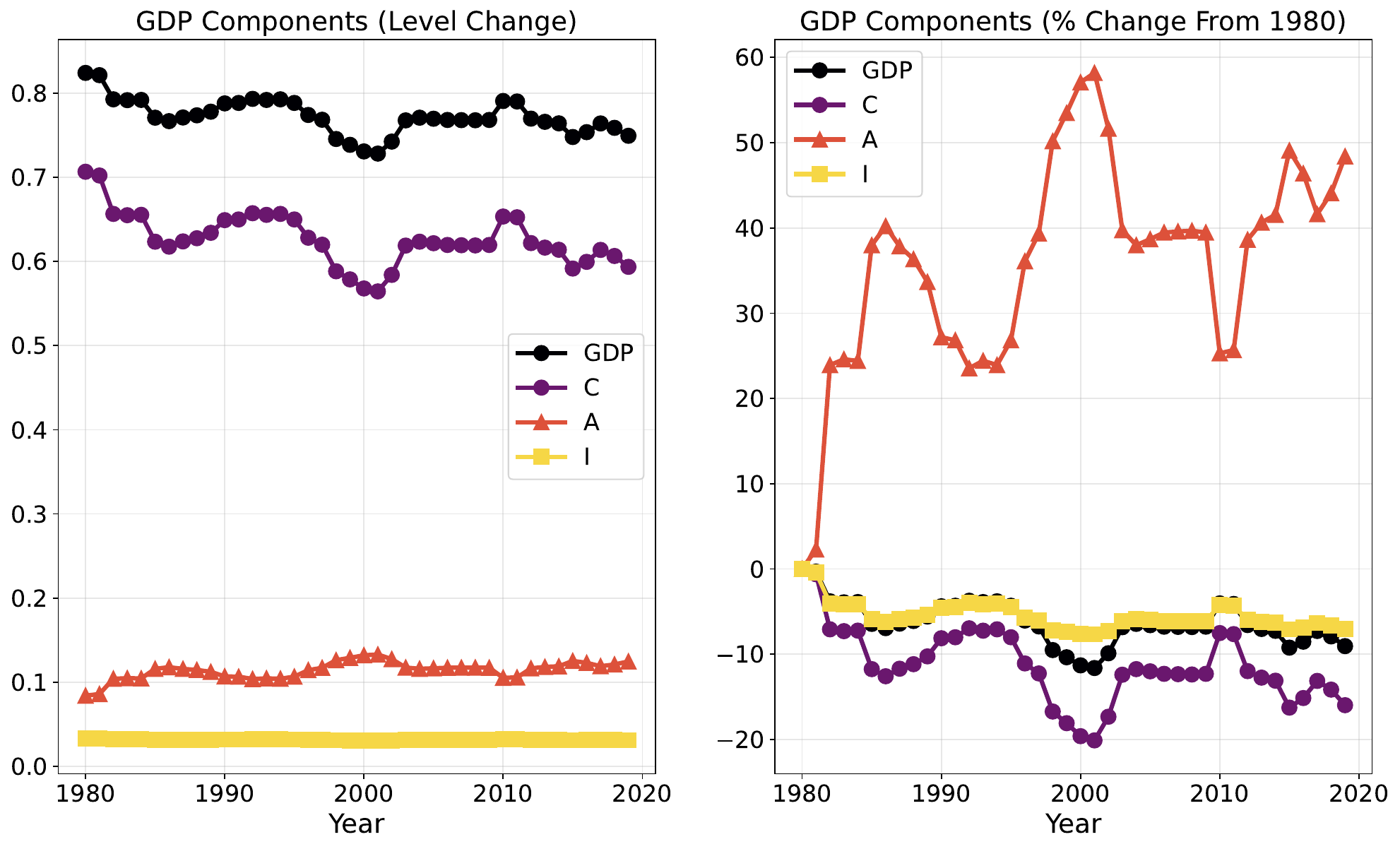}
\caption{The Response of Aggregates to the Rise in the Scale Elasticity of Demand}
\label{fig:agg_trends}
\scriptsize\raggedright{\textit{The figure plots economic aggregates, defined above, at the steady state equilibrium corresponding to each year, where we vary the the scale elasticity of demand across equilibria shown in Figure \ref{fig:phi_and_pct_neg_by_year}.}}
\end{figure}

As in Section \ref{sec:quantitative}, we consider a path of steady state equilibria corresponding to different scale elasticities of demand which are calibrated to match the rise in the percent of negative earnings between 1980 and 2019. We set the numeraire as the ideal price index, so all values are denominated in units of the household consumption good. We define investment and advertising as aggregate spending by firms on labor for capital production, $L_{k, i}$, and labor for advertising production, $L_{a,i}$. We define consumption as aggregate spending by household members on variety goods, which corresponds with the household consumption basket $C$. We define GDP as the sum of consumption, investment, and advertising.
\[
I = \int_0^1 W L_{k, i} di; \quad A = \int_0^1 W L_{a, i}; \quad C = \int_0^1 \int_0^1 P_i C_{ij} dj di; \quad \text{GDP} = C + I + A
\]
We plot each of these aggregates, in levels and in percent change from our 1980 baseline equilibrium, in Figure \ref{fig:agg_trends}. The largest percent change is in advertising, which rises by \getstat{a_pct_change}\%, while consumption and investment fall by \getstat{c_pct_change}\% and \getstat{i_pct_change}\% respectively. GDP falls by \getstat{gdp_pct_change}\%, driven by the lost consumption.

Since consumption drives the fall in GDP, and is the most welfare relevant metric, it is worth considering independently. Moreover, since the scale elasticity of demand is a parameter of the household's CES aggregator function, it is useful to disentangle its direct effect on consumption from its indirect effect via changing firm investment choices. 

\begin{figure}[!h]
\centering 
\includegraphics[width=\textwidth]{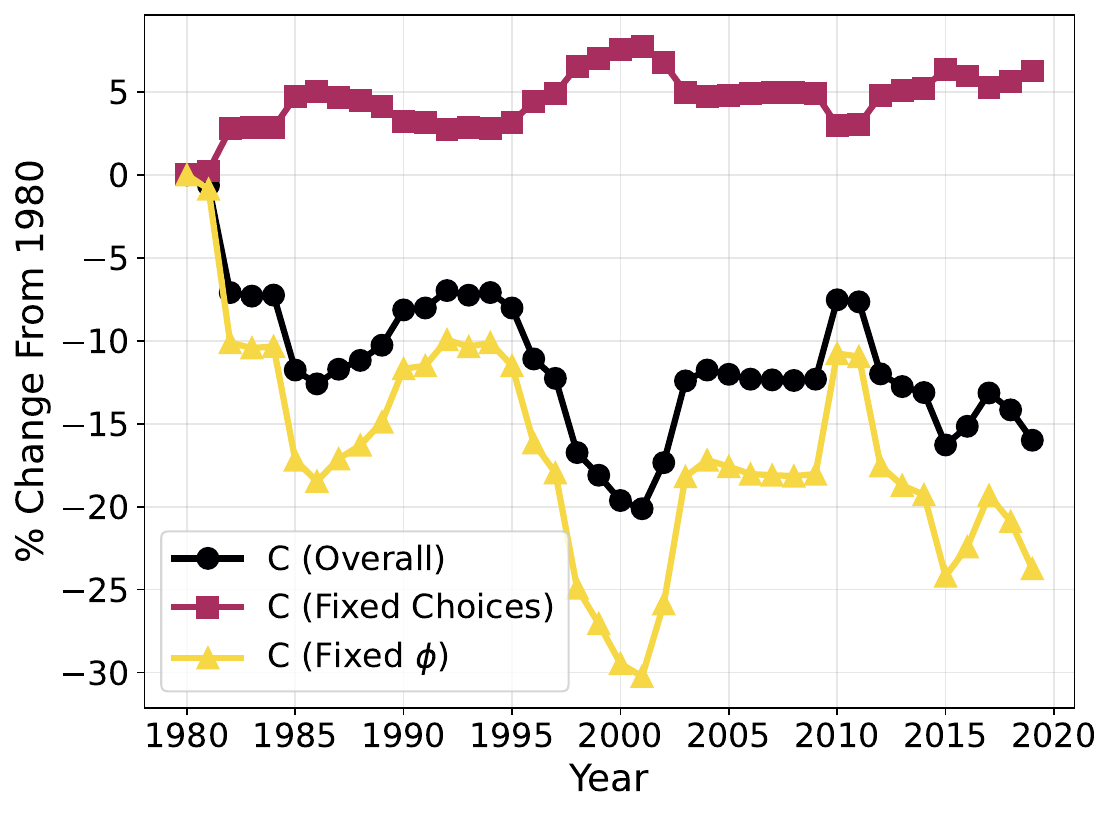}
\caption{The Direct and Indirect Effect of the Scale Elasticity on Consumption}
\label{fig:agg_c_direct_indirect}
\scriptsize\raggedright{\textit{The figure plots consumption at the steady state equilibrium corresponding to each year, alongside counterfactual consumption holding firm choices fixed while varying the scale elasticity and counterfactual consumption holding the scale elasticity fixed while varying firm choices.}}
\end{figure}

This decomposition is shown in Figure \ref{fig:agg_c_direct_indirect}: we plot consumption across equilibria, alongside (1) a counterfactual consumption path where the scale elasticity of demand changes but all firm choice variables are held fixed at the 1980 baseline equilibrium, and (2) a counterfactual consumption path where firm choice variables change but the scale elasticity of demand is held fixed at 0. The direct effect of the secular rise in the scale elasticity of demand is to \emph{raise consumption}. A formal proof of this can be found in Appendix \ref{ap:theory}; the intuition is a higher scale elasticity increases the consumption aggregator weight on varieties produced by larger firms.

\begin{figure}[!h]
\centering 
\includegraphics[width=\textwidth]{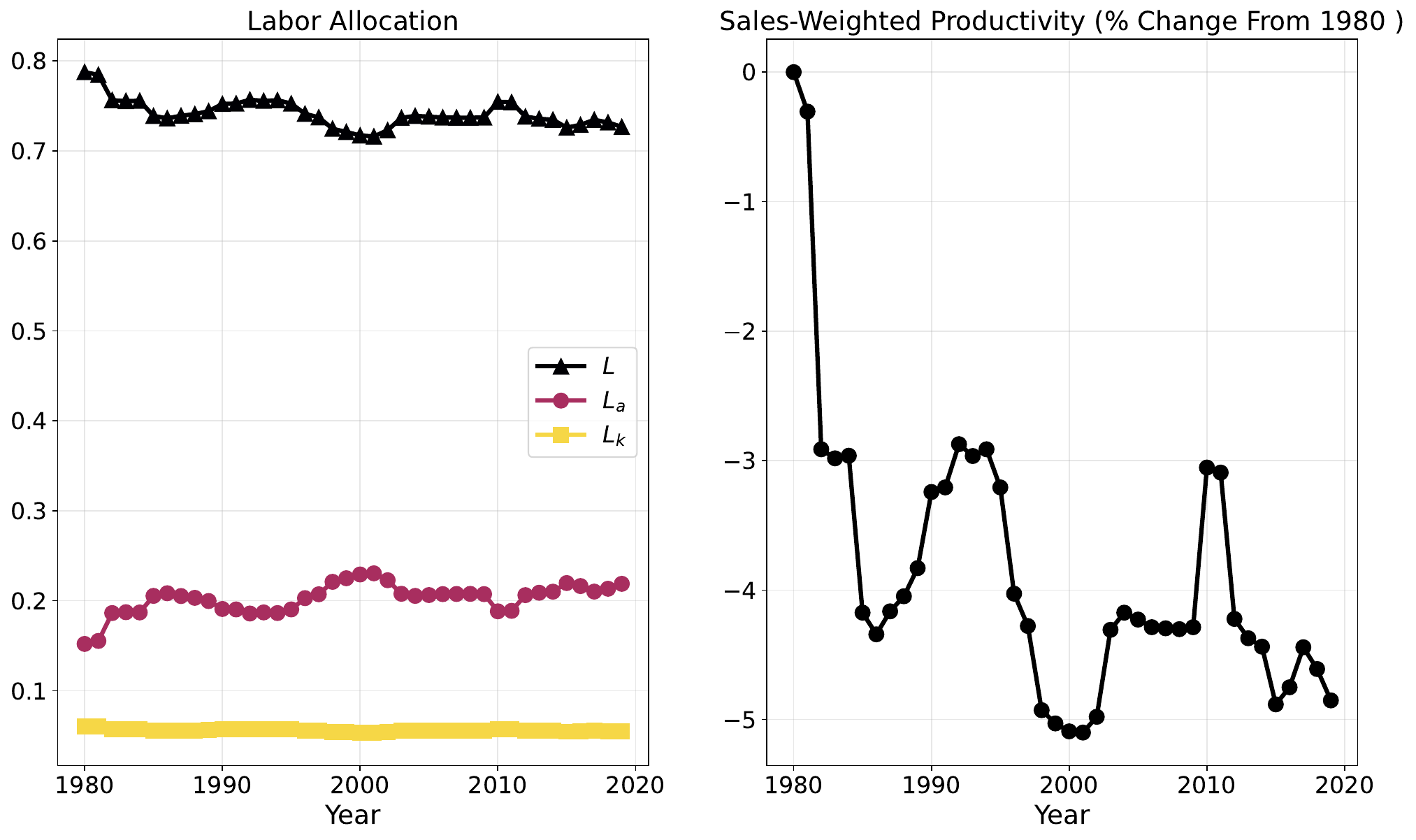}
\caption{Indirect Effect: Labor Allocation and Demand Allocation}
\label{fig:agg_c_indirect}
\scriptsize\raggedright{\textit{The figure plots the share of labor allocated to goods, advertising, and capital production (left panel) and the sales weighted productivity (right panel) at the steady state equilibrium corresponding to each year.}}
\end{figure}

The fall in consumption between 1980 and 2019 is thus driven by the indirect effect of raising the scale elasticity of demand: a higher scale elasticity changes firms' investment choices, resulting in a steady state equilibrium with lower household consumption. This is for two reasons, as shown in Figure \ref{fig:agg_c_indirect}.

First, raising the scale elasticity \emph{reallocates labor}, the sole factor of production which is in fixed supply, away from goods and capital production and towards advertising production. As discussed above, consumption is homogenous of degree zero in aggregate advertising: advertising affects relative demand, but not aggregate demand. As a consequence, any aggregate labor allocated to advertising is inherently wasteful, since any feasible allocation of demand can be sustained with the appropriate firm-to-firm relative advertising rates as overall advertising production limits to 0. As the scale elasticity rises, firms perceive greater returns on advertisements, not internalizing the aggregate implications of their decisions. They allocate a greater share of their labor to advertising production, which results in less labor being allocated to capital and goods production, compressing the production possibilities frontier of the overall economy.

Second, raising the scale elasticity \emph{reallocates demand} across variety producers. In our model, the distribution of sales across varieties depends on the demand shifting investment choices of firms. As the scale elasticity rises, the marginal value of an additional customer for new firms with small customer bases is less productivity dependent. That is, the return on an additional customer for a new, productive firm is closer to the return on an additional customer for a new, unproductive firm. This is because of the interdependence of the intensive and extensive margins of demand: as the scale elasticity $\phi$ rises, customers consume more of varieties with large customer bases, and less of varieties with small customer bases. An additional customer for a new firm with a small customer base doesn't shift the demand curve out by as much, and thus there is less of a marginal profit benefit to a lower marginal cost due to higher productivity. 

Superstar effects still exist: the marginal value on an additional customer is always higher for more productive firms, and by the inverse logic of above, it becomes more productivity dependent for incumbent firms with larger customer bases as the scale elasticity rises. However, the convex cost of advertisement and exogenous exit rate $\theta$ means the lower productivity dependence of demand shifting investment for younger firms results in a ``worse'' allocation of customers: the calibrated rise in the scale elasticity of demand between 1980 and 2019 lowers the sales weighted productivity by \getstat{sales_wtd_z_pct_change}\%.

\section*{Conclusion}

Over the last fifty years US business dynamics have changed. In this paper we document an understudied consequence of those changes: the secular rise of the percentage of firms reporting losses, both among public firms and in the broader universe of US corporations, and the increased persistence of losses year after year. We highlight that this rise has happened alongside a spreading of the sales and earnings distribution and a recomposition of firm spending away from variable costs and traditional CapEx and towards sales and general administrative expenses, which we interpret as a rise in intangible investment through sales and marketing.

We present a model to rationalize these phenomena. Our model is novel in that it incorporates \emph{demand shifting investment}, modeled as endogenous customer acquisition via advertisements, alongside traditional supply-shifting capital investment. It also includes a \emph{scale elasticity of demand}, which determines the elasticity of the intensive margin of demand (demand per customer) with respect to the extensive margin of demand (number of customers). 

This scale elasticity is unique relative to other parameters in that it has a monotonic relationship with four moments of the steady state equilibrium: the percent of firms with negative earnings, the persistence of negative earnings, the spread of the sales and earnings distribution, and the ratio of sales and marketing spending to variable costs and traditional investment. Because of this monotonic relationship, we are able to quantitatively match the rise in the percent of firms reporting losses and qualitatively match the increased persistence of losses, the spreading of the sales and earnings distribution, and the recomposition of firm spending with the scale elasticity of demand as the single driver of changes across steady state equilibria. 

The rise in the scale elasticity associated with the increase in reported losses has non-trivial aggregate implications: in our model, GDP falls due to lost consumption. Consumption falls because of the \emph{reallocation of labor} away from goods and capital production and towards advertisement production and the \emph{reallocation of demand} away from more productive firms and towards less productive firms.

\pagebreak

\bibliography{../../mylibrary.bib}

\pagebreak 

\appendix 
\numberwithin{equation}{section}
\setcounter{section}{0} \renewcommand{\thesection}{\Alph{section}}
\setcounter{equation}{0} \renewcommand{\theequation}{\Alph{section}.%
\arabic{equation}}

\section{Empirical Appendix}
\label{ap:empirical}
\setcounter{table}{0} \renewcommand{\thetable}{\Alph{section}.\arabic{table}}%
\setcounter{figure}{0} \renewcommand{\thefigure}{\Alph{section}.\arabic{figure}}%

\paragraph{Data Construction} 
The IRS measure of the percent of corporations without net income is constructed using table(s) from the IRS SOI Corporate Complete Report. The specific tables used are listed in Table \ref{tab:irs_data}, as well as the universe of returns the numbers cover. This universe is consistent from 1988-2019; pre 1988, the excluded returns don't include 1120-REIT. In general the most relevant exclusion is 1120-S (the S-Corporation). 

\paragraph{Production Function Estimation} We estimate production function parameters $\gamma_k$, $\gamma_l$, $\rho$, and $\sigma_Z$ for our Compustat sample using the methodology of \citet{ackerbergIdentificationPropertiesRecent2015}, implemented in Python by \citet{ridderHitchhikersGuideMarkup2022}. 

\paragraph{Additional Figures}

\begin{figure}[!h]
    \centering
    \includegraphics[width=\textwidth]{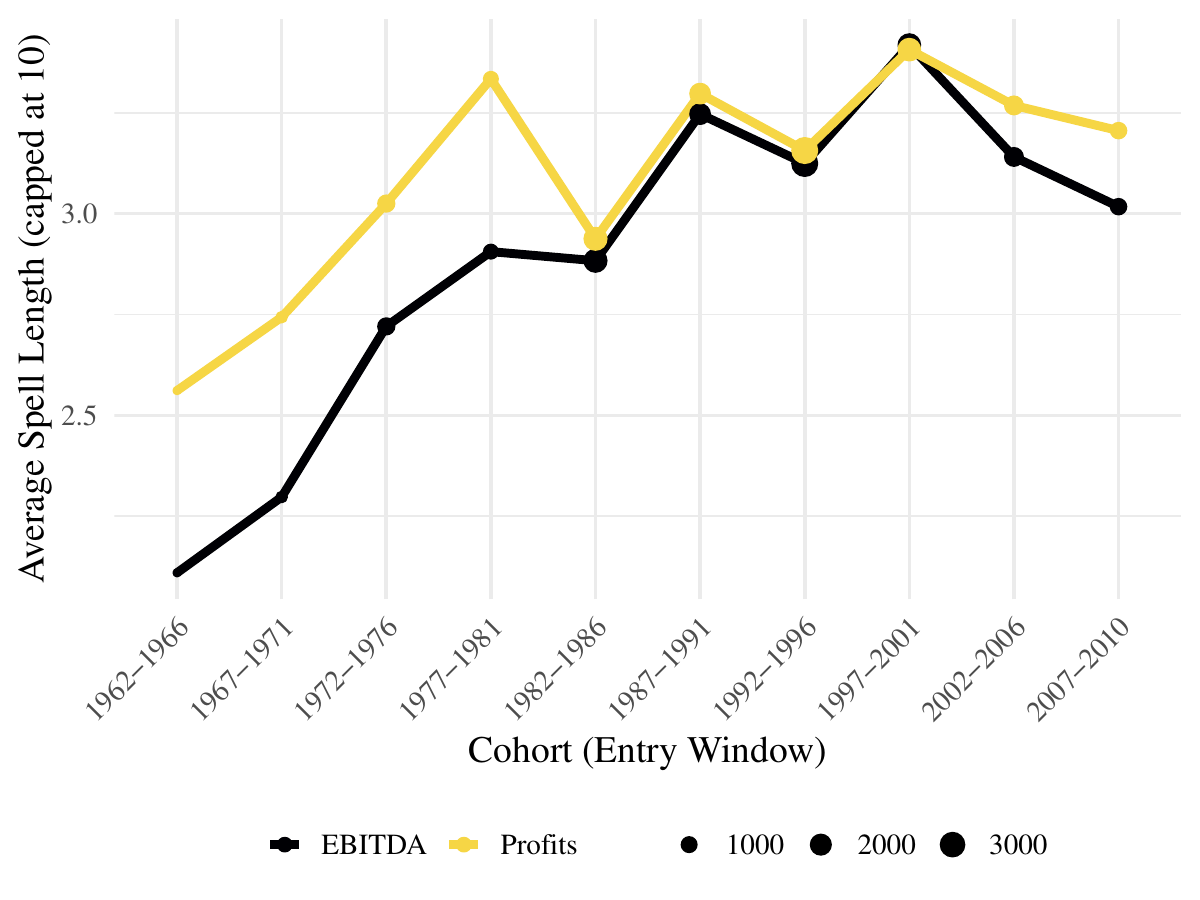}
    \caption{Increased Persistence of Losses Across Cohorts}
    \label{fig:neg_spell_cohorts}
\end{figure}

The persistence of losses (defined using EBITDA and profits) by cohort is shown in Figure \ref{fig:neg_spell_cohorts}. We also find that losses are more common among younger firms, but that the age-dependence of losses has flattened over time, as shown in Figure

\begin{figure}[!h]
    \centering
    \includegraphics[width=\textwidth]{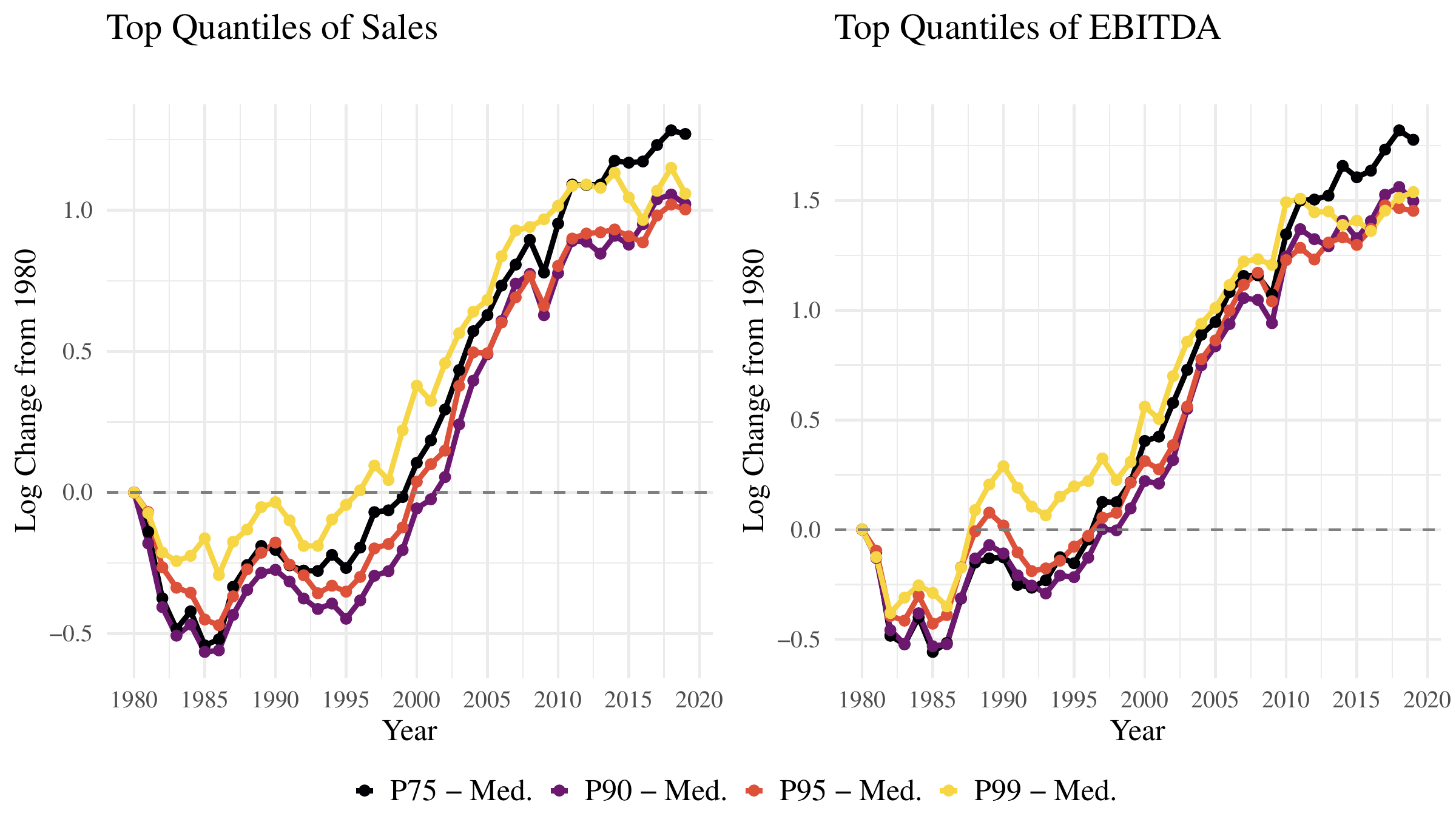}
    \caption{Top Quantiles of Sales and Earnings Over Time}
    \label{fig:top_qtiles_sales_earnings}
\end{figure}

\begin{figure}[!h]
    \centering
    \includegraphics[width=\textwidth]{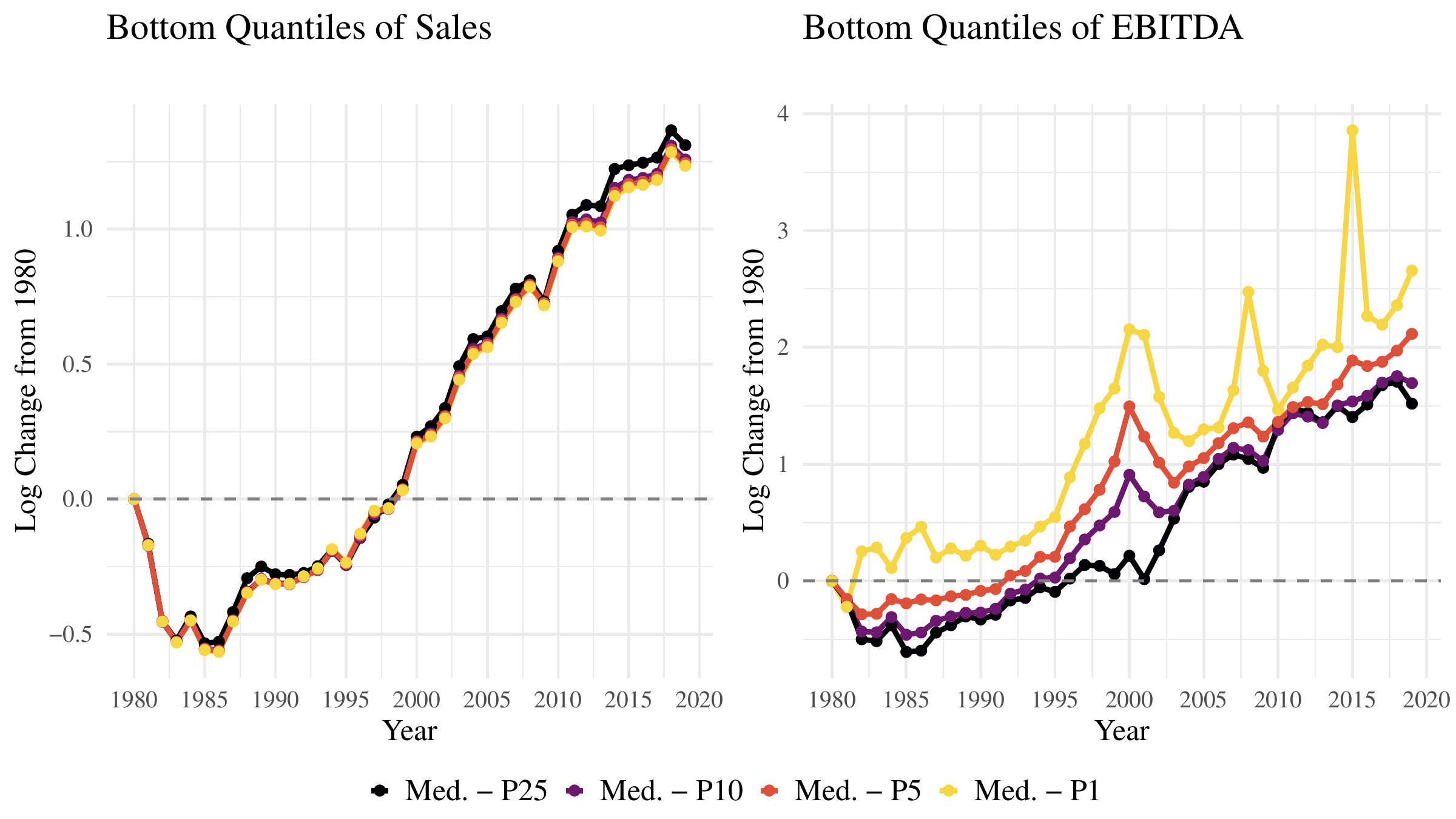}
    \caption{Bottom Quantiles of Sales and Earnings Over Time}
    \label{fig:bottom_qtiles_sales_earnings}
\end{figure}

\begin{figure}[!h]
    \centering
    \includegraphics[width=\textwidth]{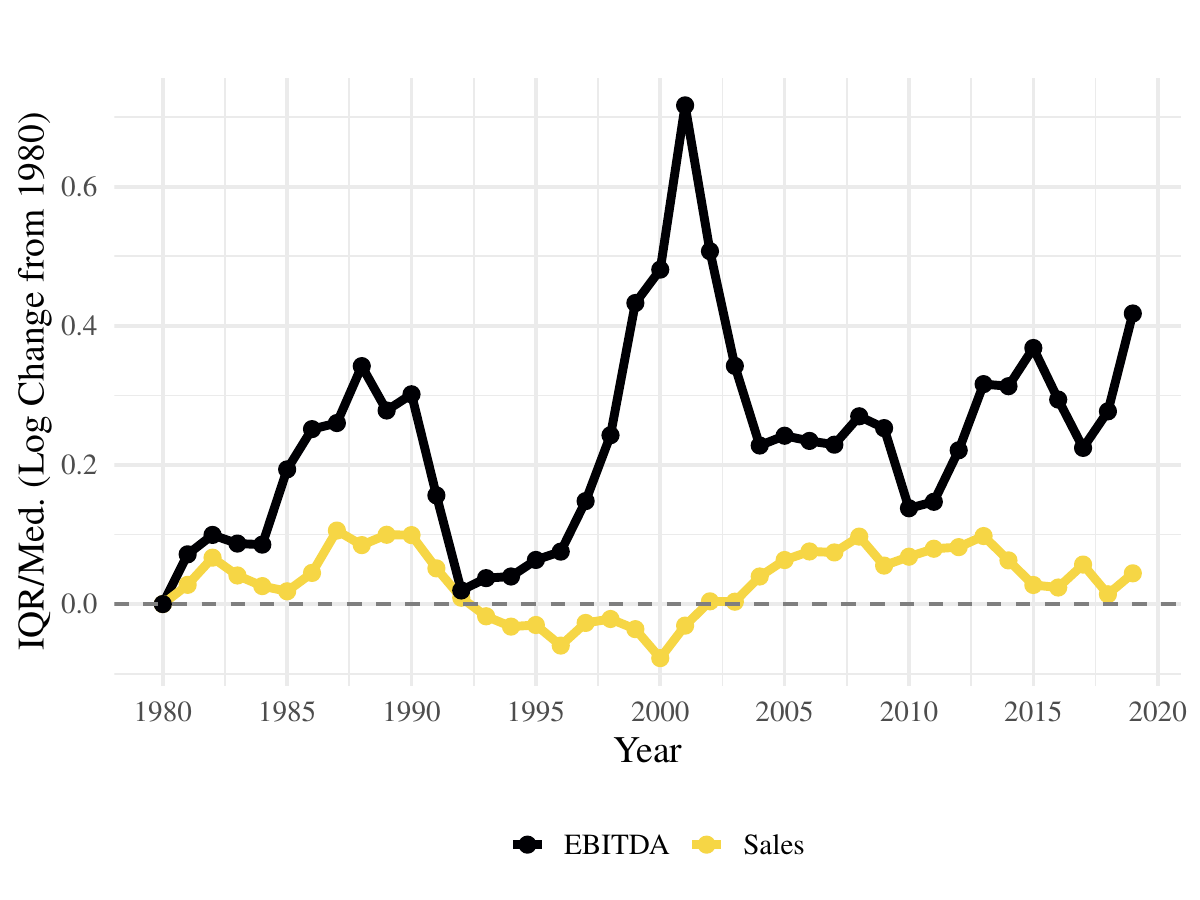}
    \caption{Scale Invariant Measures of Sales and Earnings Distribution}
    \label{fig:scale_inv_sales_earnings}
\end{figure}

The time series of quantile-level measures of spread for the sales and earnings distribution are shown in Figures \ref{fig:top_qtiles_sales_earnings} and \ref{fig:bottom_qtiles_sales_earnings}. As described in the paper, they show a very uniform spreading (all percentiles pull away from the median). Scale invariant measures of sales and earnings distribution are shown in Figure \ref{fig:scale_inv_sales_earnings}. The scale invariant measure of the earnings distribution has also risen between 1980 and 2019. The scale invariant measure of the sales distribution has not, though it is well understood that the sales distribution of the broad universe of US firms has undergone a scale invariant transformation over this period, i.e. the rise of superstar firms and sales concentration documented in \citet{autorFallLaborShare2020,kwon100YearsRising2024}.

\pagebreak 

\begin{landscape}
\begin{table}
\centering
\caption{IRS Corporate Returns without Net Income}
\label{tab:irs_data}
\scriptsize
\begin{tabular}{rlllll}
\toprule
year & Total Returns & Returns with Net Income & \% without Net Income & Table Number & Coverage \\
\midrule
1980 & 2,156,485 & 1,307,187 & 39.38\% & 17 & Returns of active corporations, other than Forms 1120S and 1120-DISC \\
1981 & 2,261,523 & 1,323,377 & 41.48\% & 16 & Returns of active corporations, other than Forms 1120S and 1120-DISC \\
1982 & 2,352,051 & 1,339,327 & 43.06\% & 16 & Returns of active corporations, other than Forms 1120S and 1120-DISC \\
1983 & 2,340,906 & 1,362,444 & 41.8\% & 16 & Returns of active corporations, other than Forms 1120S and 1120-DISC \\
1984 & 2,456,924 & 1,434,019 & 41.63\% & 16 & Returns of active corporations, other than Forms 1120S and 1120-DISC \\
1985 & 2,548,746 & 1,474,529 & 42.15\% & 16 & Returns of active corporations, other than Forms 1120S, 1120-IC-DISC, and 1120-FSC  \\
1986 & 2,597,959 & 1,507,572 & 41.97\% & 16 & Returns of active corporations, other than Forms 1120S, 1120-IC-DISC, and 1120-FSC  \\
1987 & 2,480,430 & 1,418,531 & 42.81\% & 15 & Returns of active corporations, other than Forms 1120S, 1120-IC-DISC, and 1120-FSC  \\
1988 & 2,299,896 & 1,270,952 & 44.74\% & 16 & Returns of active corporations, other than Forms 1120S, 1120-REIT, and 1120-RIC \\
1989 & 2,199,081 & 1,197,929 & 45.53\% & 16 & Returns of active corporations, other than Forms 1120S, 1120-REIT, and 1120-RIC \\
1990 & 2,136,032 & 1,131,942 & 47.01\% & 23 & Returns of active corporations, other than Forms 1120S, 1120-REIT, and 1120-RIC \\
1991 & 2,098,641 & 1,088,017 & 48.16\% & 22 & Returns of active corporations, other than Forms 1120S, 1120-REIT, and 1120-RIC \\
1992 & 2,077,518 & 1,117,309 & 46.22\% & 22 & Returns of active corporations, other than Forms 1120S, 1120-REIT, and 1120-RIC \\
1993 & 2,055,982 & 1,123,240 & 45.37\% & 22 & Returns of active corporations, other than Forms 1120S, 1120-REIT, and 1120-RIC \\
1994 & 2,310,703 & 1,272,978 & 44.91\% & 22 & Returns of active corporations, other than Forms 1120-REIT, 1120-RIC, and 1120S \\
1995 & 2,312,382 & 1,262,408 & 45.41\% & 22 & Returns of active corporations, other than Forms 1120-REIT, 1120-RIC, and 1120S \\
1996 & 2,317,886 & 1,284,278 & 44.59\% & 22 & Returns of active corporations, other than Forms 1120-REIT, 1120-RIC, and 1120S \\
1997 & 2,248,065 & 1,239,047 & 44.88\% & 22 & Returns of active corporations, other than Forms 1120-REIT, 1120-RIC, and 1120S \\
1998 & 2,249,970 & 1,239,493 & 44.91\% & 22 & Returns of active corporations, other than Forms 1120-REIT, 1120-RIC, and 1120S \\
1999 & 2,198,740 & 1,199,747 & 45.43\% & 22 & Returns of active corporations, other than Forms 1120-REIT, 1120-RIC, and 1120S \\
2000 & 2,172,705 & 1,144,496 & 47.32\% & 22 & Returns of active corporations, other than Forms 1120-REIT, 1120-RIC, and 1120S \\
2001 & 2,136,756 & 1,088,221 & 49.07\% & 22 & Returns of active corporations, other than Forms 1120-REIT, 1120-RIC, and 1120S \\
2002 & 2,100,074 & 999,680 & 52.4\% & 22 & Returns of active corporations, other than Forms 1120-REIT, 1120-RIC, and 1120S \\
2003 & 2,047,593 & 991,120 & 51.6\% & 22 & Returns of active corporations, other than Forms 1120-REIT, 1120-RIC, and 1120S \\
2004 & 2,027,613 & 1,009,794 & 50.2\% & 22 & Returns of active corporations, other than Forms 1120-REIT, 1120-RIC, and 1120S \\
2005 & 1,974,961 & 1,038,946 & 47.39\% & 22 & Returns of active corporations, other than Forms 1120-REIT, 1120-RIC, and 1120S \\
2006 & 1,955,147 & 1,008,357 & 48.43\% & 22 & Returns of active corporations, other than Forms 1120-REIT, 1120-RIC, and 1120S \\
2007 & 1,865,232 & 953,789 & 48.86\% & 22 & Returns of active corporations, other than Forms 1120-REIT, 1120-RIC, and 1120S \\
2008 & 1,782,478 & 846,540 & 52.51\% & 22 & Returns of active corporations, other than Forms 1120-REIT, 1120-RIC, and 1120S \\
2009 & 1,715,306 & 789,099 & 54\% & 22 & Returns of active corporations, other than Forms 1120-REIT, 1120-RIC, and 1120S \\
2010 & 1,671,149 & 802,991 & 51.95\% & 22 & Returns of active corporations, other than Forms 1120-REIT, 1120-RIC, and 1120S \\
2011 & 1,648,540 & 808,106 & 50.98\% & 22 & Returns of active corporations, other than Forms 1120-REIT, 1120-RIC, and 1120S \\
2012 & 1,617,739 & 825,171 & 48.99\% & 22 & Returns of active corporations, other than Forms 1120-REIT, 1120-RIC, and 1120S \\
2013 & 1,611,125 & 819,232 & 49.15\% & 22 & Returns of active corporations, other than Forms 1120-REIT, 1120-RIC, and 1120S \\
2014 & 1,601,402 & 825,267 & 48.47\% & 5.3, 5.4 & Returns of active corporations, other than Forms 1120-REIT, 1120-RIC, and 1120S \\
2015 & 1,611,236 & 806,389 & 49.95\% & 5.3, 5.4 & Returns of active corporations, other than Forms 1120-REIT, 1120-RIC, and 1120S \\
2016 & 1,574,942 & 793,047 & 49.65\% & 5.3, 5.4 & Returns of active corporations, other than Forms 1120-REIT, 1120-RIC, and 1120S \\
2017 & 1,577,412 & 785,322 & 50.21\% & 5.3, 5.4 & Returns of active corporations, other than Forms 1120-REIT, 1120-RIC, and 1120S \\
2018 & 1,546,431 & 768,223 & 50.32\% & 5.3, 5.4 & Returns of active corporations, other than Forms 1120-REIT, 1120-RIC, and 1120S \\
2019 & 1,514,347 & 737,213 & 51.32\% & 5.3, 5.4 & Returns of active corporations, other than Forms 1120-REIT, 1120-RIC, and 1120S \\
2020 & 1,489,544 & 643,544 & 56.8\% & 5.3, 5.4 & Returns of active corporations, other than Forms 1120-REIT, 1120-RIC, and 1120S \\
2021 & 1,548,569 & 714,820 & 53.84\% & 5.3, 5.4 & Returns of active corporations, other than Forms 1120-REIT, 1120-RIC, and 1120S \\
2022 & 1,556,099 & 738,736 & 52.53\% & 5.3, 5.4 & Returns of active corporations, other than Forms 1120-REIT, 1120-RIC, and 1120S \\
\bottomrule
\end{tabular}

\end{table}
\end{landscape}

\begin{table}[!h]
    \centering
    \caption{Sectoral Regressions using Sales and Marketing Spending}
    \label{tab:sector_regs}
    
\begin{tabular}{l c c c}
\hline
 & \multicolumn{3}{c}{EBITDA} \\
\cline{2-4}
 & Avg. Pct. Neg. & Avg. Neg. Spell & Std. Dev. \\
\hline
$I^{SM}$/Rev. (He et. al. 2025) & $0.75^{***}$ & $2.67^{***}$ & $3.31^{***}$ \\
                                & $(0.04)$     & $(0.15)$     & $(0.43)$     \\
\hline
Num. obs.                       & $2459$       & $2459$       & $2459$       \\
\hline
\multicolumn{4}{l}{\scriptsize{$^{***}p<0.001$; $^{**}p<0.01$; $^{*}p<0.05$}}
\end{tabular}

\end{table}

\begin{figure}[!h]
    \centering
    \includegraphics[width=\textwidth]{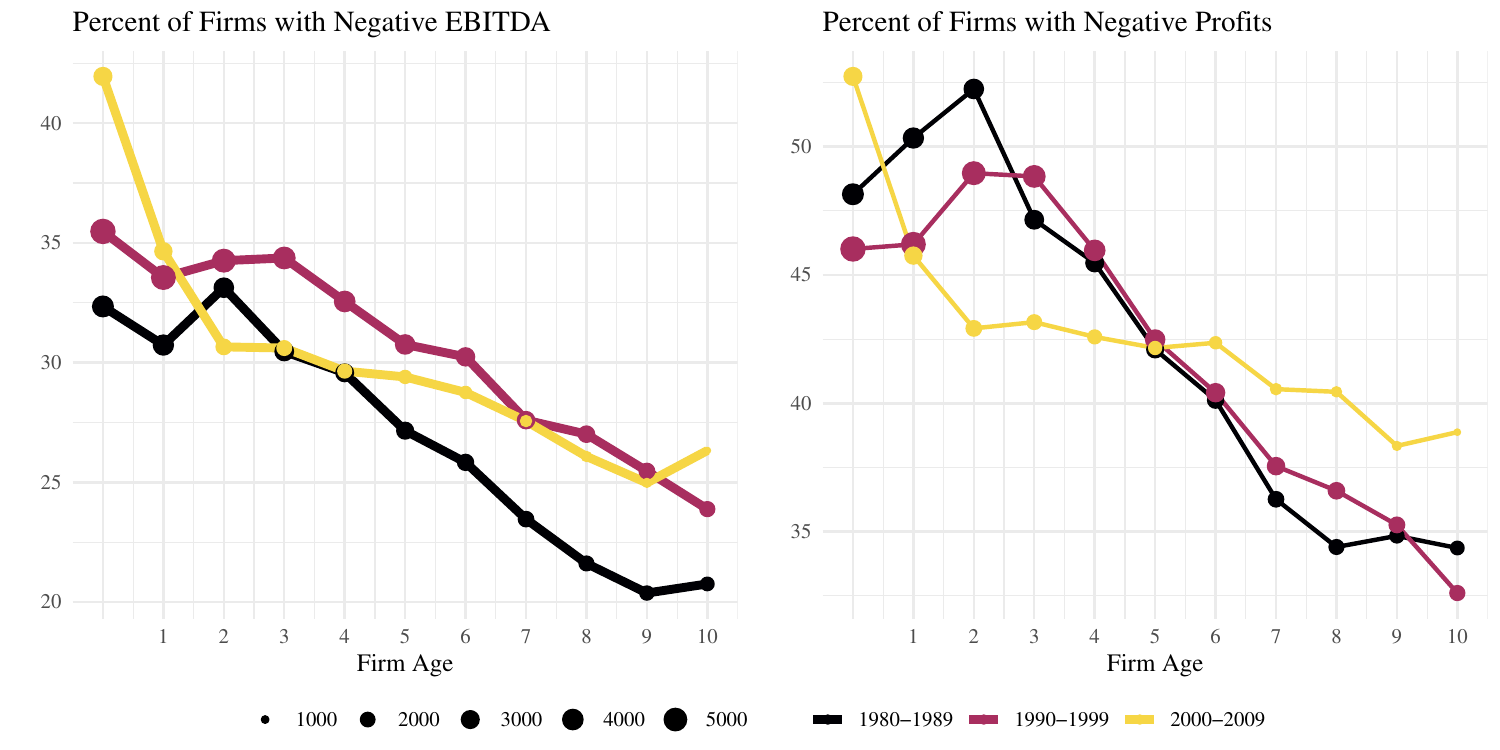}
    \caption{Percent of Firms with Negative Earnings, by Age and Cohort}
    \label{fig:neg_earnings_ageANDcohorts}
\end{figure}

\begin{figure}[!h]
    \centering
    \includegraphics[width=\textwidth]{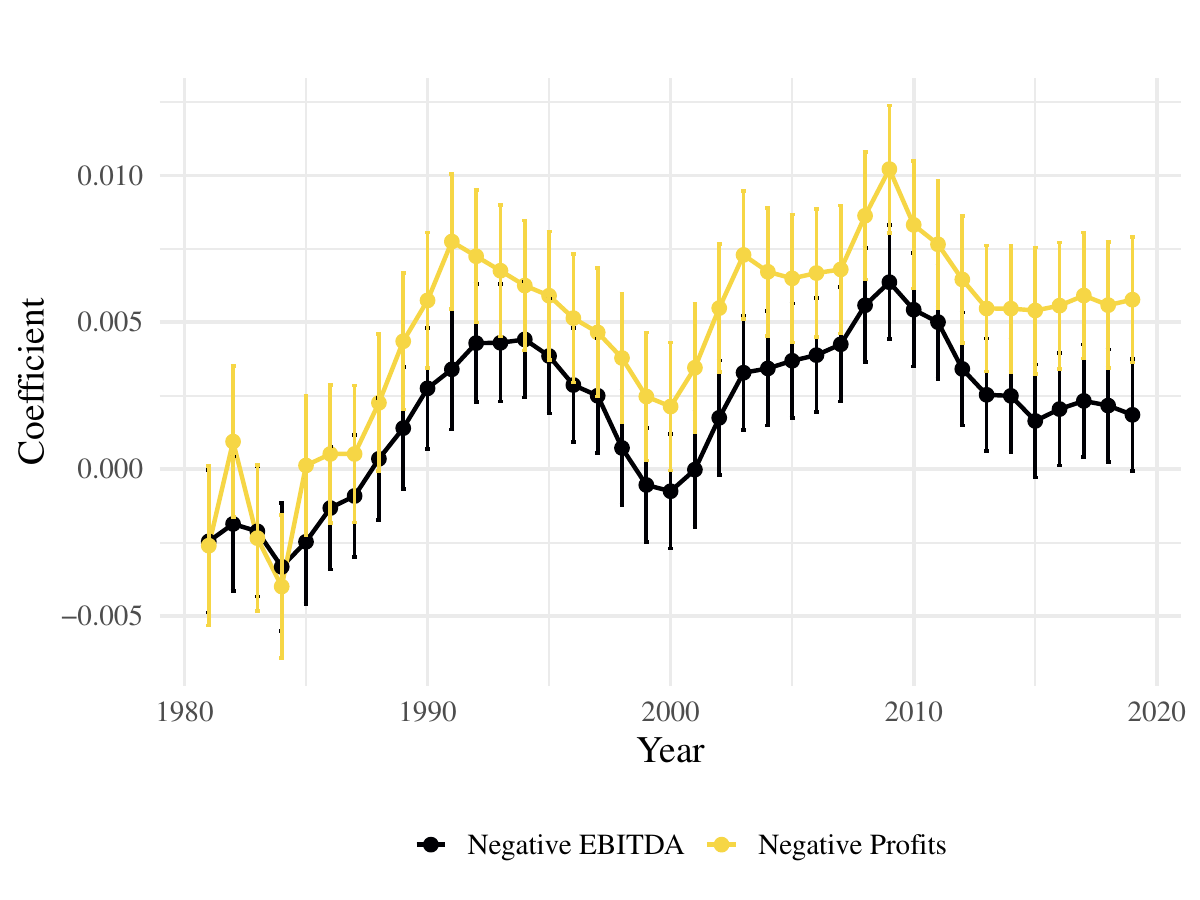}
    \caption{Percent of Firms with Negative Earnings, by Age and Cohort}
    \label{fig:neg_earnings_age_coefs}
\end{figure}

\clearpage

\section{Theoretical Appendix}
\label{ap:theory}
\setcounter{table}{0} \renewcommand{\thetable}{\Alph{section}.\arabic{table}}%
\setcounter{figure}{0} \renewcommand{\thefigure}{\Alph{section}.\arabic{figure}}%

\paragraph{Deriving Profit Elasticities}
From the household problem, given customer base $M_{it}$, the firm's demand function is...
\[
Y_{it} = M_{it}^{1+\phi} P_{it}^{-\sigma} C_t \rightarrow P_{it} = M_{it}^{\frac{1+\phi}\sigma} C_t^{1/\sigma} Y_{it}^{-1/\sigma}
\]
Assuming Cobb-Douglas production, our static cost function is wage times labor 
\[
C(Y_i) = W\left(\frac{Y_i}{Z_i K_i^{\gamma_k}}\right)^{\frac{1}{\gamma_l}} \rightarrow C'(Y_i) = W (Z_i K_i^{\gamma_k})^{\frac{-1}{\gamma_l}} \frac{1}{\gamma_l} Y_i^{\frac {1 - \gamma_l} {\gamma_l}}
\]
Plugging into monopolist price setting condition we get...
\begin{align*}
(M_{it}^{1+\phi} C_t)^{1/\sigma} Y_{it}^{-1/\sigma} & = W_t (Z_{it} K_{it}^{\gamma_k})^{\frac{-1}{\gamma_l}} \frac{1}{\gamma_l} Y_{it}^{\frac {1 - \gamma_l} {\gamma_l}}\left(\frac{\sigma}{\sigma-1} \right) \\
Y_i^{\frac {1 - \gamma_l} {\gamma_l} + \frac 1 \sigma} & = \gamma_l (M_i^{1+\phi} C)^{\frac 1\sigma} (Z_i K_i^{\gamma_k})^{\frac 1 {\gamma_l}}W^{-1} \left(\frac{\sigma-1}{\sigma}\right) \\
Y_i(M_i, K_i, Z_i) & = \left(\gamma_l (M_i^{1+\phi} C)^{\frac 1\sigma} (Z_i K_i^{\gamma_k})^{\frac 1 {\gamma_l}}W^{-1}\left(1 - \frac 1 \sigma\right)\right)^{\frac{\sigma \gamma_l}{\sigma (1 - \gamma_l) + \gamma_l}}
\end{align*}
Plugging into the definition of profits we get
\[
\pi(M_i,K_i, Z_i) = \frac{\sigma}{\sigma-1} C'(Y_i) Y_i - C(Y_i) = C(Y_i) \left(\frac{\mu}{\gamma_l} - 1\right) 
\]
Plugging in the definition of output in to the cost function we get
\begin{align*}
C(Y_i) & =  W\left(Z_i K_i^{\gamma_k}\right)^{\frac{-1}{\gamma_l}}\left(\gamma_l (M_i^{1+\phi} C)^{\frac 1\sigma} (Z_i K_i^{\gamma_k})^{\frac 1 {\gamma_l}}W^{-1}\left(1 - \frac 1 \sigma\right)\right)^{\frac{\sigma}{\sigma (1 - \gamma_l) + \gamma_l}} \\
& = \left(W^{\frac 1 \Lambda - 1} (Z_i K_i^{\gamma_k})^{\frac 1 {\gamma_l}(1 - \frac 1 \Lambda)} (M_i^{1+\phi} C)^{\frac 1 \sigma} \frac{\gamma_l}{\mu} \right)^\Lambda \\
\pi(M_i, K_i, Z_i) & = \left(W^{\frac 1 \Lambda - 1} (Z_i K_i^{\gamma_k})^{\frac 1 {\gamma_l}(1 - \frac 1 \Lambda)} (M_i^{1+\phi} C)^{\frac 1 \sigma} \frac{\gamma_l}{\mu} \right)^\Lambda\left(\frac{\mu}{\gamma_l} - 1 \right) \\
R(M_i, K_i, Z_i) & = \left(W^{\frac 1 \Lambda - 1} (Z_i K_i^{\gamma_k})^{\frac 1 {\gamma_l}(1 - \frac 1 \Lambda)} (M_i^{1+\phi} C)^{\frac 1 \sigma} \frac{\gamma_l}{\mu} \right)^\Lambda\left(\frac{\mu}{\gamma_l} \right)
\end{align*}
Where $\Lambda = \frac{\sigma}{(1-\gamma_l) \sigma + \gamma_l}$, $\mu = \frac{\sigma}{\sigma - 1}$. This delivers profit elasticities...
\begin{align*}
\epsilon^\pi_K & = \frac{\gamma_k (\Lambda - 1)}{\gamma_l} = \frac{\gamma_k(\sigma - (1-\gamma_l)\sigma - \gamma_l)}{\gamma_l ((1-\gamma_l) \sigma + \gamma_l)} = \frac{\gamma_k(\sigma - 1 )}{(1-\gamma_l) \sigma + \gamma_l} \\
\epsilon^\pi_M & = \frac {1+\phi} \sigma \Lambda = \frac {1+\phi} {(1-\gamma_l) \sigma + \gamma_l}
\end{align*}

\paragraph{Effect of Scale Elasticity on Consumption Aggregator}
Formal derivation TK

\paragraph{Effect of Scale Elasticity on $\frac{\partial \pi}{\partial M_i \partial Z_i}$}
Formal derivation TK

\pagebreak

\section{Quantitative Appendix}
\setcounter{table}{0} \renewcommand{\thetable}{\Alph{section}.\arabic{table}}%
\setcounter{figure}{0} \renewcommand{\thefigure}{\Alph{section}.\arabic{figure}}%

\subsection{Solving the Firm Problem}

We solve for the marginal value functions $V_M(M,K,Z) \equiv \partial V/\partial M$ and $V_K(M,K,Z) \equiv \partial V/\partial K$. By the envelope theorem these satisfy
\begin{align}
  V_M(M,K,Z) &= \pi_M(M,K,Z) + \beta(1-\theta)(1-\delta_m)\,\mathbb{E}_Z[V_M(M',K',Z')], \label{eq:env_m}\\
  V_K(M,K,Z) &= \pi_K(M,K,Z) + \beta(1-\theta)(1-\delta_k)\,\mathbb{E}_Z[V_K(M',K',Z')]. \label{eq:env_k}
\end{align}
The algorithm is initialized with the guess $V_M^{(0)} = \pi_M$ and $V_K^{(0)} = \pi_K$ and iterates equations \eqref{eq:env_m}--\eqref{eq:env_k} to convergence.

The first-order conditions for advertising and capital investment are
\begin{align}
  W P_M\, (L^a)'(a) &= \beta(1-\theta)\,\mathbb{E}_Z[V_M(M',K',Z')], \label{eq:foc_a}\\
  W\, (L^k)'(i) &= \beta(1-\theta)\,\mathbb{E}_Z[V_K(M',K',Z')]. \label{eq:foc_k}
\end{align}
Following the endogenous grid method (EGM) we invert \eqref{eq:foc_a} and \eqref{eq:foc_k}  analytically. The marginal investment-cost functions are
\[
  (L^a)'(a) = \frac{1}{\alpha_a} a^{1/\alpha_a - 1}, \qquad (L^k)'(i) = \frac{1}{\alpha_k} i^{1/\alpha_k - 1},
\]
with closed-form inverses
\[
  [(L^a)']^{-1}(x) = \left(\alpha_a\,x\right)^{\alpha_a/(1-\alpha_a)}, \qquad
  [(L^k)']^{-1}(x) = \left(\alpha_k\,x\right)^{\alpha_k/(1-\alpha_k)}.
\]

Because each firm has \emph{two} continuous endogenous states, we alternate 1-D EGM: we decouple the two investment decisions by fixing one state on its exogenous grid while solving for the other endogenously (this is in practice faster than doing both at once because we avoid a 2-D interpolation step).

\paragraph{Pass A (customer-capital lines).}
For each fixed $(K'_j, Z_s)$ on the next-period capital grid and productivity grid:
\begin{enumerate}
  \item Evaluate $\Lambda^a_{j,s}(M') \equiv \beta(1-\theta)\,\mathbb{E}_{Z'}[V_M(M', K'_j, Z')\mid Z_s]$ on the $M'$ grid.
  \item Invert the FOC \eqref{eq:foc_a}: $a(M') = [(L^a)']^{-1}(\Lambda^a_{j,s}(M') / (W P_M))$, clipped to $[0, M_{\max} P_M]$.
  \item Recover implied current $M$: $M = [M' - a(M')/P_M]/(1-\delta_m)$. 
  \item Invert $M \mapsto M'$ via linear interpolation to obtain $M'(M)$ on the structured grid.
  \item Store $\Lambda^a_{j,s}(M'(M))$ on the structured $(M, K'_j, Z_s)$ grid.
\end{enumerate}

\paragraph{Pass B (capital lines).}
Symmetrically, for each fixed $(M'_i, Z_s)$:
\begin{enumerate}
  \item Evaluate $\Lambda^k_{i,s}(K') \equiv \beta(1-\theta)\,\mathbb{E}_{Z'}[V_K(M'_i, K', Z')\mid Z_s]$ on the $K'$ grid.
  \item Invert the FOC \eqref{eq:foc_k}: $i(K') = [(L^k)']^{-1}(\Lambda^k_{i,s}(K') / W)$, clipped to $[0, K_{\max}]$.
  \item Recover implied current $K = [K' - i(K')]/(1-\delta_k)$.
  \item Invert the monotone map to obtain $K'(K)$ on the structured grid.
  \item Store $\Lambda^k_{i,s}(K'(K))$ on the structured $(M'_i, K, Z_s)$ grid.
\end{enumerate}

\paragraph{Policy and envelope updates.}
After both passes, the on-grid marginal continuation values $\hat{\Lambda}^a$ and $\hat{\Lambda}^k$ are used to recover policies via the FOC inverses on the structured $(M,K,Z)$ grid. The marginal value functions are then updated via the envelope conditions \eqref{eq:env_m}--\eqref{eq:env_k}:
\begin{align*}
  V_M^{\text{new}}(M,K,Z) &= \pi_M(M,K,Z) + (1-\delta_m)\,\hat{\Lambda}^a(M,K,Z)\cdot W P_M,\\
  V_K^{\text{new}}(M,K,Z) &= \pi_K(M,K,Z) + (1-\delta_k)\,\hat{\Lambda}^k(M,K,Z)\cdot W.
\end{align*}
We iterate until convergence of marginal value functions.

\subsection{Additional Figures}

\begin{figure}[!h]
    \centering
    \includegraphics[width=.9\textwidth]{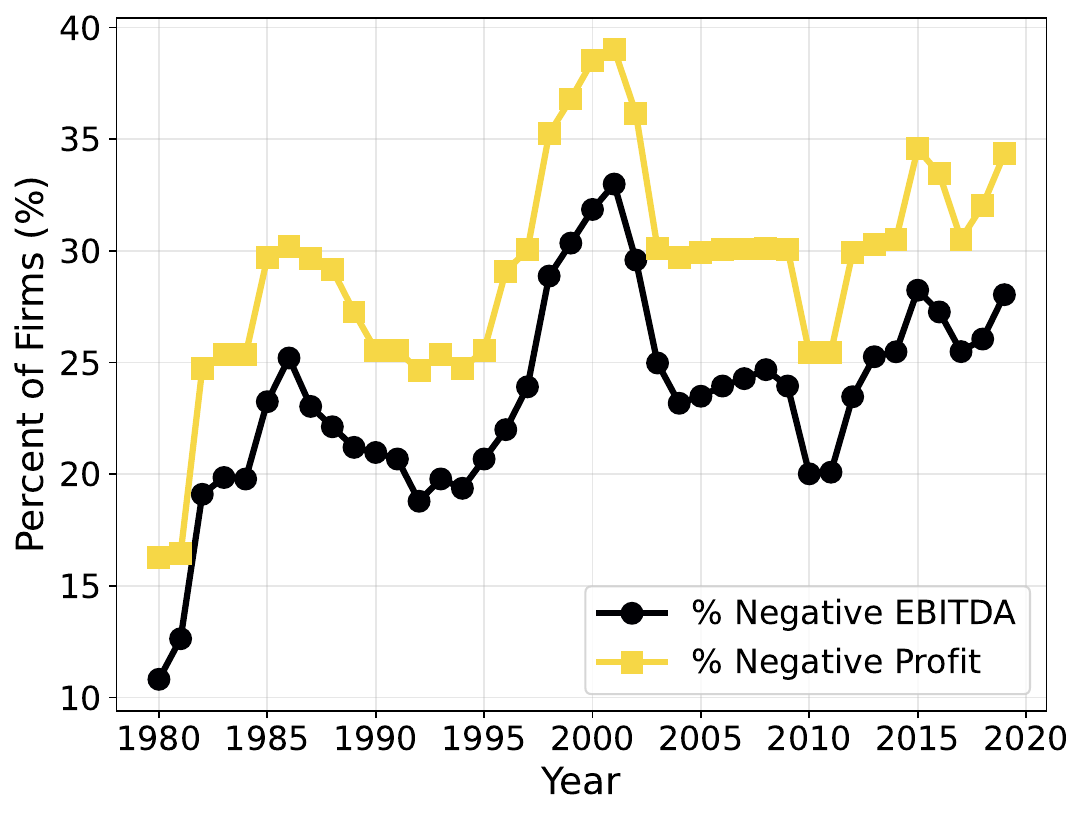}
    \caption{Negative Profits and Negative Earnings in Model}
    \label{fig:neg_ebitda_neg_profits}
\end{figure}

\begin{figure}[!h]
    \centering
    \includegraphics[width=.9\textwidth]{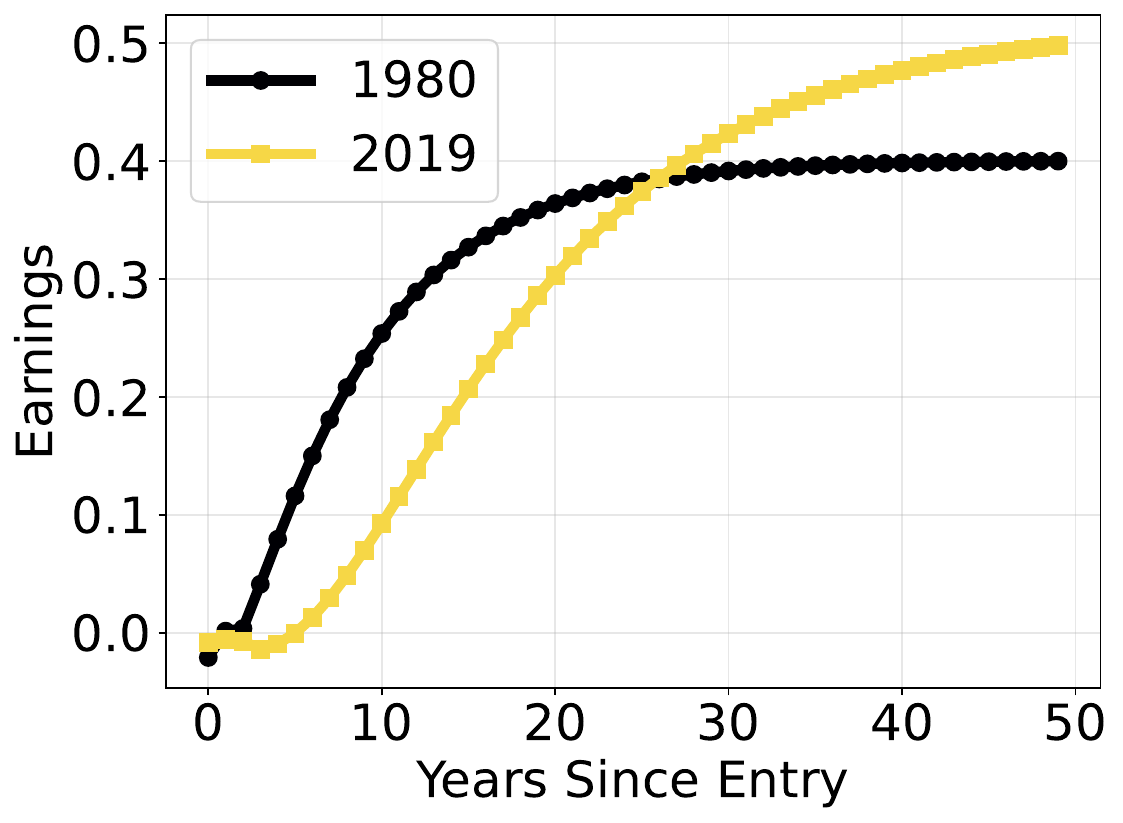}
    \caption{Earnings Path for Median Productive Firm, 1980 vs. 2019}
    \label{fig:earning_path_1980_2019}
\end{figure}

\end{document}